\documentclass[english]{article}
\usepackage[T1]{fontenc}
\usepackage{babel}
\usepackage[left=3.5cm,top=3.5cm,right=3cm,bottom=3cm]{geometry}
\pdfoutput=1
\usepackage{float}
\usepackage{graphicx}  
\usepackage{subfigure} 
\usepackage{amsmath}
\usepackage{amssymb}
\usepackage[labelformat=parens,labelsep=quad,skip=3pt]{caption}

\usepackage{setspace}
\usepackage{lineno}
\usepackage{multirow}
\usepackage{hyperref}
\usepackage{soul}
\usepackage[vlined,ruled]{algorithm2e}
\hypersetup{
    colorlinks=true,
    linkcolor=blue,
    filecolor=magenta,      
    urlcolor=cyan,
}
\makeatletter

\usepackage{beamerarticle,pgf}

\newcommand\makebeamertitle{\frame{\maketitle}}
\AtBeginDocument{
	\let\origtableofcontents=\tableofcontents
	\def\tableofcontents{\@ifnextchar[{\origtableofcontents}{\gobbletableofcontents}}
	\def\gobbletableofcontents#1{\origtableofcontents}
}

\makeatother

\newcommand{\cF}{\mathcal{F}}

\newcommand{\A}{\mathbf{A}}
\newcommand{\Y}{\mathbf{Y}}
\newcommand{\y}{\mathbf{y}}
\newcommand{\D}{\mathbf{D}}
\newcommand{\dm}{\mathbf{d}}
\newcommand{\T}{\mathbf{T}}
\newcommand{\X}{\mathbf{X}}
\newcommand{\x}{\mathbf{x}}

\newcommand{\um}{\mathbf{u}}
\newcommand{\XD}{\mathbf{X} \mathbf{D}}
\newcommand{\W}{\mathbf{W}}
\newcommand{\C}{\mathbf{C}}
\newcommand{\SIGMA}{\boldsymbol{\Sigma}}
\newcommand{\MU}{\boldsymbol{\mu}}
\newcommand{\BETA}{\boldsymbol{\beta}}
\newcommand{\XI}{\boldsymbol{\xi}}
\newcommand{\Epsi}{\boldsymbol{\varepsilon}}

\usepackage{soul} 
\usepackage{color}
\definecolor{morado}{RGB}{125,0,177}
\definecolor{rojo}{RGB}{255,0,0}
\definecolor{azul}{RGB}{0,0,255}
\definecolor{verde}{RGB}{0,135,0}

\newcommand{\cris}{\color{azul} }  
\newcommand{\sirc}{ \color{black}} 
\newcommand{\stcris}[1]{\cris \st{#1} \sirc} 

\newcommand{\laot}{40} 
\newcommand{\latt}{45} 
\newcommand{\lao}{40} 
\newcommand{\lat}{55} 

\usepackage[toc,page]{appendix}
\usepackage{bbm} 

\begin{document}
\title{Uncertainty Quantification for Fault Slip Inversion}
\author{J Cricelio Montesinos-L\'opez \footnotemark[1]\ \footnotemark[4]
	\and Antonio Capella \footnotemark[2]
	\and J Andr\'es Christen \footnotemark[1]
	\and Josu\'e Tago \footnotemark[3]   
}
\renewcommand{\thefootnote}{\fnsymbol{footnote}}

\footnotetext[1]{Centro de Investigaci\'on en Matem\'aticas (CIMAT), Jalisco S/N, Valenciana, Guanajuato, 36023, M\'exico.
\textit{jose.montesinos, jac at cimat.mx}}
\footnotetext[2]{Instituto de Matem\'aticas, Universidad Nacional Aut\'onoma de M\'exico, M\'exico City, M\'exico.
\textit{capella@im.unam.mx}}
\footnotetext[3]{Facultad de Ingenier\'ia, Universidad Nacional Aut\'onoma de M\'exico, M\'exico City, M\'exico.
 \textit{josue.tago@gmail.com}}

\footnotetext[4]{Corresponding author}
\makebeamertitle

We propose an efficient Bayesian approach to infer a fault displacement from geodetic data in a slow slip event. Our physical model of the slip process reduces to a multiple linear regression subject to constraints. Assuming a Gaussian model for the geodetic data and considering a multivariate truncated normal prior distribution for the unknown fault slip, the resulting posterior distribution is also multivariate truncated normal. Regarding the posterior, we propose an algorithm based on Optimal Directional Gibbs that allows us to efficiently sample from the resulting high-dimensional posterior distribution of along dip and along strike movements of our fault grid division. A synthetic fault slip example illustrates the flexibility and accuracy of the proposed approach. The methodology is also applied to a real data set, for the 2006 Guerrero, Mexico, Slow Slip Event, where the objective is to recover the fault slip on a known interface that produces displacements observed at ground geodetic stations. As a by-product of our approach, we are able to estimate moment magnitude for the 2006 Guerrero Event with uncertainty quantification. 

\section{Introduction} \label{sec:1}

A major task of geophysics is to make quantitative statements about Earth's interior in terms of surface measurements.  One fundamental element of earthquake investigations is to estimate the magnitude and distribution of slip along a fault plane. Fault slips may consist of complex and heterogeneous source processes, while limited geodetic data typically leads to an ill-posed inverse problem (IP). Conventionally, regularization is used to transform such IPs into a well-posed optimization problem for a single-source model. The most common approach is to add Tikhonov regularization terms to smooth the solution \cite{Calvetti_2000, mccaffrey2007fault, wallace2010diverse, radiguet2011spatial}, as well as including positivity constraints and reducing the solution space \cite{Tago_2020}.  All these regularization terms are supported by the physical processes being modeled. These strategies make IPs solvable and computationally feasible. However, the solution uncertainty evaluation is not methodologically found solely using regularization.  These schemes produce only limited point-wise solution estimates, such as the maximum a posteriori (MAP). Moreover, the lack of sufficient physical interpretation of some critical regularization terms may introduce bias in the solutions without justification. For more robust and informative IP solutions, and to formally quantify their uncertainty, we require a different approach.
 
The Bayesian statistical approach provides a rigorous framework to handle constrain and uncertainty quantification (UQ) of IPs. A likelihood model for observations is assumed, and prior information is incorporated using probability density functions (pdf) to determine the posterior distribution through Bayes theorem, which quantifies our inference's uncertainty. A prior pdf is established through its interpretation as a modeling device of the probabilistic prior knowledge available on source movements and imposing model restrictions using truncated pdfs \cite{fukuda2008fully, minson2013bayesian,amey2018bayesian,Nocquet_2018}.

Bayesian techniques have been of limited use for slip inversions, mainly because the simulation from the posterior distribution is not straightforward. This typically occurs when there are many parameters, such as in Slow Slip Events (SSE, see below for more details).

In modern Bayesian analyses, Markov Chain Monte Carlo (MCMC) algorithms \cite{robert2013monte} are standard tools to sample from the posterior distribution. Many versions of the MCMC method have been proposed in the literature, but the Metropolis-Hastings (MH) and the Gibbs sampler algorithms are the most common \cite{robert2013monte}. 


\cite{minson2013bayesian} developed a framework for Bayesian inversion of finite fault earthquake models. They combined a Metropolis algorithm with simulated annealing and genetic algorithms to sample high-dimensional problems in a parallel computing framework. The method remains computationally expensive despite parallelization. \cite{maatouk2017gaussian} proposes a finite-dimensional Gaussian processes approximation to allow for inequality constraints in the entire domain. Their initial problem is equivalent to simulate a Gaussian vector restricted to convex sets, and they use an improved rejection sampling \cite{maatouk2016new} in which only the random coefficients in the convex set are selected. They mention that the multivariate truncated normal (MTN) simulation can be accelerated by MCMC methods or Gibbs sampling \cite{geweke1991efficient}.


There are many methods available for simulating the MTN distribution \cite{BRESLAW19941,756335,robert1995simulation,yu2011efficient}. Most of these methods are based on the Gibbs sampler, which is simple to use and has the advantage of accepting all proposals generated and, therefore, is not affected by poor acceptance rates, such as rejection sampling. These methods work well in many situations but may be very slow if we have a high correlation and high dimensionality. Also, few of these methods have been applied for estimating uncertainty in geostatistical inverse modeling.

\cite{Nocquet_2018} shows that a MTN prior can be applied to achieve positivity or bound constraints. He employs recent findings in MTN probability calculations \cite{genz2009computation} to derive relevant posterior statistics (e.g., posterior marginal pdf, mean and covariances) without performing MCMC sampling. However, the evaluation of these quantities require complex numerical integration over an hyper-rectangle while quantifying the uncertainty of a function $f$ of the parameters (e.g., moment magnitude $M_w$) is not straightforward. In contrast, Monte Carlo samples when evaluated on $f$, the posterior uncertainty of the quantity of interest can be directly obtained. \cite{michalak2008gibbs} provides a statistically rigorous methodology for geostatistical interpolation and inverse modeling, subject to multiple and spatially-variable inequality constraints. The approach uses a Gibbs sampler to characterize the marginal probability distribution at each estimation point, using a MTN prior probability distribution. This kind of algorithms are systematic Gibbs samplers which makes CPUtime increase linearly with dimension \cite{christen2017optimal}. \cite{christen2017optimal} explore an optimality criterion for the MCMC Direction Gibbs algorithm to simulate from a MTN distribution. This criterion consists of minimizing the Mutual Information between two consecutive steps of the Markov chain. The algorithm proposed in \cite{christen2017optimal} is especially suited for high correlation and high dimensionality; one of the main advantages is that CPU time per iteration does not increase linearly with dimensionality.



In this work, we propose an efficient Bayesian approach for estimating the parameters in a constrained multiple linear regression model. Combining \cite{christen2017optimal} and \cite{Cricelio2016}, we propose an Optimal Directional Gibbs algorithm that allows us to sample from high-dimensional problems efficiently when the posterior distribution is a MTN distribution. Besides presenting a synthetic example, we apply our method to quantify the uncertainty in the IP of seismic slip along the subduction interface in the 2006 Guerrero, Mexico, SSE. Moreover, with our method, we are able to provide the posterior distribution of the moment magnitude for this event.

A SSE is a slip produced at a fault that do not generate seismic waves. However, the induced deformation may be registered at the surface from SSEs lasting several weeks to a couple of months. SSEs have been observed in different fault configurations around the world \cite{Gao_et_al_2012_BSSA}, and the role they play in the seismic cycle is an active research topic \cite{Kato_et_al_2012_Science, Ruiz_et_al_2014_Science, Radiguet_2016, Cruz-Atienza_2020}.

In Mexico, SSEs have been identified in different segments along the subduction region, on the Pacific coast, where the Cocos Plate and the North American Plate collide. In the so-called Guerrero GAP (GGap), before the great Mw8.2 Tehuantepec event on 8 September 2017, SSEs showed a periodicity of approximately four years and a duration from six to twelve months \cite{Cruz-Atienza_2020}.
The SSE, occurred within Guerrero state in 2006, one of the most studied globally, was recorded at 15 continuous GPS stations \cite{radiguet2011spatial, Radiguet_2012, Cavalie_2013, Bekaert_2015, Tago_2020}. This event offers the opportunity to analyze the slip's spatial evolution and delimit the characteristics of a typical SSE in the GGap. In this case, the IP consists of recovering the slip along a known interface that produces displacements at the surface observed at 15 continuous GPS geodetic stations.


\section{Methodology} \label{sec:Method}

\subsection{Bayes inference}\label{sec:BUQ}

A wide range of applications are concerned with the solution of an IP \cite{kaipio2011bayesian}: given some observations of an output, $\y = (y_1, \ldots, y_n)$, determine the corresponding inputs $\theta$  such that $y_i = \cF (\theta) + ~\text{error}$.
We refer to the evaluation of $\cF$ as solving the forward problem, and consequently, $\cF$ is called the Forward Map (FM). In general, the FM is a complex non-linear map, with input parameters  $\theta$, defined by an initial/boundary value problem for a system of ordinary or partial differential equations.

IPs appear in many branches of science and mathematics, mainly in situations where quantities of interest are different from those we can measure.  In IPs, model parameter values must be estimated from the observed data.  Until recently, an IP has been widely accepted as a statistical problem. With IPs there may be no solution, or the solution may not be unique or it may depend sensitively on measurements $y_i$ \cite{kaipio2006statistical}. A way to approach these difficulties is to formulate the IP in the Bayesian framework. \cite{stuart2010inverse} studied conditions for the well-posedness of the Bayesian formulation of IPs. In this framework, a noise model is assumed for the observations, for example, an additive Gaussian noise model
$$
y_i = \cF (\theta) + \varepsilon_i,
$$
where the errors, $\varepsilon_i$, follow a normal distribution with mean zero and variance $\sigma^2$.
This observational model generates a probability density of $\y$ given the parameters $\theta$, namely $P_{\Y |\Phi} \left( \y| \theta,\sigma \right)$.  For fixed data $\y$, and as a function of $\theta$, we call this the likelihood function. Based on the available information, a prior model $P_{\Phi}(\cdot)$ is stated for $\Phi =(\theta,\sigma)$, and a posterior distribution is obtained through Bayes theorem
$$
P_{\Phi | \Y}\left(\theta, \sigma| \y \right)  = \frac{P_{\Y | \Phi} \left(\y | \theta,\sigma \right) P_{\Phi} \left(\theta, \sigma \right)} {P_\Y \left( \y \right)}.
$$
The denominator is the normalization constant, also called the marginal likelihood of the observations or model evidence. 

In a frequentist (classical) statistical paradigm, we often maximize the likelihood to obtain a single estimate for the parameter of interest. Uncertainty is defined by the sampling distribution based on the idea of infinite repeated sampling. 
In contrast, the goal of Bayesian inversion is not only to obtain a single estimate for the unknowns but to quantify their uncertainty consistently with the observed data. Therefore, we describe the unknowns by probability distributions. Before any observation is available, we have much uncertainty in the unknown. After making the measurements, the uncertainty is reduced, and the task is to quantify it and provide probabilistic answers to questions of interest \cite{kaipio2011bayesian}. Previous information regarding the physics of the problem, which is not specific enough to be incorporated into the direct problem, may be incorporated into the prior probability distribution.

\subsection{Forward map} \label{sec:FM}
For the direct problem, we begin with the representation theorem for the elastostatic equations which models the displacement $\um ( \x )$, at the coordinates $\x$ of the GPS station, due to a slip $\dm (\XI)$, produced at a fault $\SIGMA$, as 
\begin{equation}
u_{j} \left( \x \right) = \int_{\Sigma} T_{k} \left( S_{ij} \left( \XI ; \x \right),\hat{\mathbf{n}} \left( \XI \right) \right) d_{k} \left( \XI \right) d\SIGMA, \qquad j\in\left\{ x,y,z\right\} \label{eq:RepTeo}
\end{equation}
where $T_{k} \left( \cdot, \cdot \right)$ is the $k$-component of the traction on the fault computed through the Somigliana tensor, $S_{ij} \left( \XI; \x \right)$, and the fault normal vector $\hat{\mathbf{n}} \left( \XI\right)$ \cite{udias2014}. If the traction and the slip are projected along the dip component, $d$-direction, and along the strike direction, $s$-direction, Eq.~(\ref{eq:RepTeo}) can be written in matrix form as
\begin{align*}
\left[ \begin{array}{c}
u_{x}\left(\text{\ensuremath{\x}}\right)\\
u_{y}\left(\text{\ensuremath{\x}}\right)\\
u_{z}\left(\text{\ensuremath{\x}}\right)
\end{array} \right] & = \int_{\SIGMA} \begin{bmatrix}
T_{s}\left(S_{ix}\left(\XI;\x\right),\hat{\mathbf{n}}\left(\XI\right)\right) & T_{d}\left(S_{ix}\left(\XI;\x\right),\hat{\mathbf{n}}\left(\XI\right)\right)\\
T_{s}\left(S_{iy}\left(\XI;\x\right),\hat{\mathbf{n}}\left(\XI\right)\right) & T_{d}\left(S_{iy}\left(\XI;\x\right),\hat{\mathbf{n}}\left(\XI\right)\right)\\
T_{s}\left(S_{iz}\left(\XI;\x\right),\hat{\mathbf{n}}\left(\XI\right)\right) & T_{d}\left(S_{iz}\left(\XI;\x\right),\hat{\mathbf{n}}\left(\XI\right)\right)
\end{bmatrix}\begin{bmatrix}d_{s}\left(\XI\right)\\
d_{d}\left(\XI\right)
\end{bmatrix}d\SIGMA,
\end{align*}
or in a more compact vector notation as
\begin{equation*}
\um \left(\text{\ensuremath{\x}}\right)  
= \int_{\SIGMA}  \T \left(\XI; \x \right) \dm \left( \XI \right) d\SIGMA .
\end{equation*}

We assume to know the fault's geometry, which is discretized in $M$ subfaults, $\{\XI^{1}, \XI^{2}, \ldots, \XI^{M} \}$, such that the integral can be approximated as
$$
\um \left( \text{\ensuremath{\x}} \right) \simeq\sum_{i=1}^{M}A^{i} \T \left( \XI^{i}; \x \right) \dm \left( \XI^{i} \right),
$$
where $A^{i}$ is the $i$-subfault area. Finally, if we want to compute the displacement for $N$ receivers, we can order the displacements in a single vector such that the entire computation is reduced to a simple matrix-vector product as
\begin{align*}
\left[\begin{array}{c}
\um \left( \x^{1} \right)\\
\um \left( \x^{2} \right)\\
\vdots\\
\um \left( \x^{N} \right)
\end{array}\right] & =\begin{bmatrix}A^{1}\T\left(\XI^{1};\x^{1}\right) & A^{2}\T\left(\XI^{2};\x^{1}\right) & \cdots & A^{M}\T\left(\XI^{M};\x^{1}\right)\\
A^{1}\T\left(\XI^{1};\x^{2}\right) & A^{2}\T\left(\XI^{2};\x^{2}\right) & \cdots & A^{M}\T\left(\XI^{M};\x^{2}\right)\\
\vdots & \vdots & \ddots & \vdots\\
A^{1}\T\left(\XI^{1};\x^{N}\right) & A^{2}\T\left(\XI^{2};\x^{N}\right) & \cdots & A^{M}\T\left(\XI^{M};\x^{N}\right)
\end{bmatrix}\begin{bmatrix}\dm\left(\XI^{1}\right)\\
\dm\left(\XI^{2}\right)\\
\vdots\\
\dm\left(\XI^{M}\right)
\end{bmatrix}, \nonumber 
\end{align*}
or more compactly as
\begin{equation}
\mathbf{U}  = \XD, \label{eq:FLM}
\end{equation}
where $\mathbf{U} \in \mathbb{R}^{3N}$, $\X \in \mathbb{R}^{3N \times 2M}$, and $\D \in \mathbb{R}^{2M}$. 

\subsection{Data likelihood} \label{sec:likelihood}
The IP consists of recovering the slip at each subfault, of a known interface, that produces displacements observed at geodetic stations. Due to the linearity of the FM in Eq.~\eqref{eq:FLM}, we solve the Bayesian inversion as a multiple linear regression model with constraints on the coefficients.
We use a simple representation of observation and modeling errors by assuming a Gaussian multiple linear model
\begin{equation*}
\Y  = \XD + \Epsi,
\end{equation*}
where $\Epsi$ follows a Gaussian distribution, $\Epsi \sim N_{3N} \left( \mathbf{0}, \Sigma \right)$, and $\Sigma = \mathbf{I} \otimes \gamma$ is a known covariance matrix of observation errors. That is,
\begin{equation}\label{eq:Model}
\mathbf{Y}|\D \sim N_{3N} \left(\XD, \Sigma \right),
\end{equation}
where $\mathbf{Y} \in \mathbb{R}^{3N}$ are the displacements observed at the $N$ geodetic stations stored in a single ordered vector, as in Eq.~\eqref{eq:FLM}. $\Sigma$ is a covariance matrix of the misfits between the observations and our predictions: $\gamma = diag \left( \left[ \sigma_{x}^{2}, \sigma_{y}^{2}, \sigma_{z}^{2} \right] \right)$ are the North, East, and Vertical deviations, $\mathbf{I}$ is an identity matrix of order $N$, and $\otimes$ denote the Kronecker product.
Therefore, the likelihood is given by
\begin{equation}\label{eq:Likelihood}
\pi \left( \mathbf{Y} |\D \right) = \left( 2 \pi \right) ^{-3N/2} \left| \A \right| ^ {1/2} \exp\left\{ -\frac{1}{2} \left(\mathbf{Y} - \X \D \right)^{T} \A \left( \mathbf{Y} - \X \D \right) \right\} ,
\end{equation}
where $\A = \Sigma^{-1}$ is the precision matrix.

\subsection{Prior elicitation} \label{sec:Prior}

Bayesian formulation of IPs requires that we specify a prior distribution for each model parameter. Proposal of the prior density is an essential step of Bayesian analyses and is often the most challenging and critical part of the approach. Usually, the major problem while proposing an adequate prior density lies in the nature of the prior information.
The prior specification is less critical for large sample sizes since the likelihood typically dominates the posterior distribution. The prior distribution plays a much more crucial role in small sample sizes because the posterior distribution represents a compromise of the prior knowledge and the observed evidence. For the 2006 Guerrero SSE, there are only $45$ observations and more than a thousand parameters to estimate (for the proposed model), so defining an adequate prior distribution is crucial for the inversion.  On the other hand, the problem is so ill posed that the inversion becomes useless unless prior knowledge is put into the problem in terms of at least simple restrictions on the possible solutions for $\D$. Thus, the importance of including such prior knowledge within the framework of a formal Bayesian approach. 

\subsubsection{Gaussian process priors}

In statistics, a Gaussian process (GP) is a stochastic process (a collection of random variables indexed by time or space), such that every finite collection of those random variables has a multivariate normal distribution.

The most commonly used probability densities in statistical IPs are undoubtedly Gaussian since they are easy to construct. However, they form a much more versatile class of densities than is usually believed \cite{kaipio2006statistical}. 

For the slip vector $\D = (d_s^1,d_d^1,d_s^2,d_d^2,\ldots, d_s^M, d_d^M)^T$, we consider a GP prior distribution, that is, $\D \sim N\left( \mathbf{0}, \frac{1}{\sigma_{\beta}^2}\A_{0} \right)$, but with truncated support, $d_{s}^i \in (a_s, b_s )$ and $d_{d}^i \in (a_d, b_d )$, $i = 1, \ldots , M$, where $\frac{1}{\sigma_{\beta}^2}\A_{0}$ denote the precision matrix, with $\A_{0} = \BETA \W \C^{-1} \W \BETA$ and $\sigma_{\beta}^2$ is an unknown scale factor that characterizes the magnitude of $\D$. Appendix \ref{Ape:Sigmas}, explains how to configure this hyperparameter, and $\BETA = \mathbf{I} \otimes \underline{\beta}$, with $\underline{\beta} = diag\left(\left[\beta_{s}, \beta_{d}\right]\right)$, where different precisions, $\beta_{s}$ and $\beta_{d}$, are considered for the along strike and along dip components, respectively. We consider an along dip variance five times greater than the along strike variance(i.e., $\beta_{s} = 1$ and $\beta_{d} = 1/5$); since we expect most of the slip along the opposite of the subduction direction. The matrix $\W$ of weights were included in the inversion scheme to penalize movements at depths greater than $z_{lim}=50$ km
\begin{align}
\W\left(i,j\right) & =\begin{cases}
1+0.5\left(\text{depth}\left(i,j\right)-z_{lim}\right)/1e3 & \text{depth}\left(i,j\right)>z_{lim}\\
1 & \text{depth}\left(i,j\right)\leq z_{lim}.
\end{cases}\label{eq:Wei}
\end{align}
The correlation matrix $\C$ is used to introduce correlation between parameters of nearby subfaults, and it is constructed using the Mat\'ern covariance function, explained in Sect. \ref{Sec:Mater}. Thus, the prior density is 
\begin{equation}
\pi \left( \D \right) = \frac{1}{Z_{prior}}  \exp\left\{ -\frac{1}{2 \sigma_{\beta}^2} \D^{T}\A_{0}\D \right\} \mathbbm{1}_{ (\mathbf{a},\mathbf{b})}(\D),
\label{eq:PriorD}
\end{equation}
where $\mathbbm{1}_{(\mathbf{a},\mathbf{b})}$ is the indicator function, $\mathbf{a} = \mathbf{1}_M \otimes [a_s,a_d]$, $\mathbf{b} = \mathbf{1}_M \otimes [b_s,b_d]$, with $\mathbf{1}_M$ the all-ones vector of length $M$, and $Z_{prior}$ is an unknown normalization constant of this MTN distribution. The constraints $a_s \leq d_{s}^i \leq b_s$, and $a_d \leq d_{d}^i \leq b_d$, $i = 1, \ldots , M$, imposed on $\D$ are based on prior information on the physical processes being modeled. If we assume that the coupling has been removed from the GPS data, then it should only consider the displacement due to an SSE. However, we allow some negative slip ($a_d <0 $) for the dip component since, on one hand, we expect that there will be subfaults where there is no slip\footnote{If a random variable is positive, then its expected value is positive. In this way, if we consider $a_d > 0$, we would be forcing small slips in all subfaults.}, on the other hand, the coupling removal is not precise. 
Together with the constraints on the support, the density function Eq.~\eqref{eq:PriorD} is indeed a MTN distribution.

\subsubsection{Mat\'ern covariance}\label{Sec:Mater}
The Mat\'ern covariance \cite{minasny2005} is a covariance function widely used in spatial statistics to define the covariance between measurements made at two points separated by $d$ distance units. 
In a GP, the essential ingredient is the covariance function, and this is used to introduce a correlation between nearby points (i.e., spatial smoothing). To construct the correlation matrix $\C$, we use the most simplified form of the Mat\'ern covariance function corresponding to $C ^1$ function (i.e., the space of functions that admit derivatives of first order). In the discretized subfaults model, the element $(i, j)$ of $\C$ is given by the relation
\begin{equation}
\C \left( i, j \right) = \gamma^{2} \left( 1 + \sqrt{3} \frac{d\left( i, j \right)}{\lambda} \right) \exp \left\{ -\sqrt{3}\frac{d \left( i, j \right)}{\lambda} \right\} ,
\label{eq:Corr}
\end{equation}
where $d(i,j)$ is the distance between the subfault $i$ and the subfault $j$, $\gamma^{2}$ represent the variance, and $\lambda$ is the correlation length. Note that as $\lambda$ increases, more coefficients of the matrix $\C$ become relevant (i.e., more subfaults are correlated).


Different correlation lengths, $\lambda_{s}$ and $\lambda_{d}$, are considered for the along -strike and along -dip components, respectively. Also, we consider that the strike component of subfault $i$ has zero correlation with the dip component of the subfault $j$.  That is, we assume independence between the strike and the dip components since, by construction, the dip component is perpendicular to the strike component. The optimal correlation lengths are chosen to minimize the Deviance Information Criterion (DIC); see Sect~\ref{sec:Deviance} for more details.

We consider $\gamma^{2}=1$ in Eq.~(\ref{eq:Corr}) to control the variance of the slips with the scale factor $\sigma_{\beta}^2$, the weight matrix $\W$ and the vector $\underline{\beta} = diag\left(\left[\beta_{s}, \beta_{d}\right]\right)$, throughout the precision matrix, $\A_{0} = \frac{1}{\sigma_{\beta}^2} \BETA \W \C^{-1} \W \BETA$, of the density function given in Eq.~(\ref{eq:PriorD}).

\subsection{Posterior distribution} \label{sec:Posterior}

The likelihood function given in Eq.~(\ref{eq:Likelihood}) and the prior distribution given in Eq.~(\ref{eq:PriorD}) are combined via Bayes' theorem to form the so-called \textit{posterior distribution}, namely
\begin{align*}
\pi\left( \D| \mathbf{Y} \right) & \propto \pi \left( \mathbf{Y} |\D \right) \pi \left( \D \right) \\
 & \propto \exp \left\{ -\frac{1}{2} \left( \mathbf{Y} - \X \D \right)^{T} \A \left(\mathbf{Y} - \XD \right) \right\} \exp \left\{ -\frac{1}{2\sigma_{\beta}^2} \D^{T}\A_{0} \D\right\} \mathbbm{1}_{ (\mathbf{a},\mathbf{b})}(\D) \\
 & \propto \exp \left\{ -\frac{1}{2} \left[ \D^{T} \X^{T} \A \X \D - 2 \mathbf{Y}^{T} \A\XD + \frac{1}{\sigma_{\beta}^2} \D^{T} \A_{0} \D \right] \right\} \mathbbm{1}_{ (\mathbf{a},\mathbf{b})}(\D)\\
 & = \exp\left\{ -\frac{1}{2} \left[ \D^{T} \left(\X^{T} \A\X + \frac{1}{\sigma_{\beta}^2}\A_{0} \right)\D - 2 \D^{T} \X^{T} \A \mathbf{Y} \right] \right\} \mathbbm{1}_{ (\mathbf{a},\mathbf{b})}(\D)\\
 & =\exp\left\{ -\frac{1}{2} \left[ \D^{T}\A_{p}\D - 2 \D^{T} \A_{p} \A_{p}^{-1} \X^{T}\A\mathbf{Y} \right] \right\} \mathbbm{1}_{ (\mathbf{a},\mathbf{b})}(\D)\\
 & = \exp\left\{ -\frac{1}{2} \left[ \left( \D - \MU_{p}\right)^{T}\A_{p}\left(\D - \MU_{p} \right) \right]\right\} \mathbbm{1}_{ (\mathbf{a},\mathbf{b})}(\D),
\end{align*}
where $\A_{p} = \X^{T}\A\X + \frac{1}{\sigma_{\beta}^2}\A_{0}$, $\A_{0} = \BETA \W \C^{-1} \W \BETA$ and $\MU_{p} = \A_{p}^{-1} \X^{T} \A \mathbf{Y}$. Thus,
\begin{align*}
\pi\left( \D| \mathbf{Y} \right) &  = \frac{1}{Z_{post}} \exp\left\{ -\frac{1}{2} \left[ \left( \D - \MU_{p}\right)^{T}\A_{p}\left(\D - \MU_{p} \right) \right]\right\} \mathbbm{1}_{ (\mathbf{a},\mathbf{b})}(\D),
\end{align*}
where $Z_{post}$ is an unknown normalization constant. Therefore, $\D|\Y$ has a MTN distribution.

To obtain information from $\D|\mathbf{Y}$ one needs to calculate relevant posterior statistics, e.g., marginals of the subfaults movements, expected values, quantifying the uncertainty of a function of $\D$, etc. This is not trivial and accordingly we propose to use a MCMC sampler. In the next section, we propose an Algorithm to simulate from the MTN distribution.
\subsection{Posterior exploration and MCMC} \label{sec:Exploration}
The MCMC simulation methods are algorithms used to produce samples from a $\pi$ distribution, which is usually complex, without simulating such distribution directly. These methods are based on constructing an Ergodic Markov chain $\mathbf{X}^{(t)}$ whose stationary distribution is precisely $\pi$. These methods have proven to be very useful in several areas, particularly in Bayesian Statistics \cite{minson2013bayesian,maatouk2017gaussian,michalak2008gibbs}.

The Gibbs sampler \cite{gelfand1990sampling} is an MCMC algorithm that, systematically or randomly, simulates conditional distributions on a set of directions. A general case of the Gibbs sampler is the Optimal Direction Gibbs sampling, which chooses an arbitrary direction $\boldsymbol{e}\in\mathbb{R}^{n}$ such that $\left\Vert \boldsymbol{e}\right\Vert =1$, and sampling from the conditional distribution along such direction\footnote{Note that if we take the directions set $\mathbf{e}$ as the canonical directions and are chosen systematically, the standard Gibbs sampler is obtained, whereas the canonical directions are taken randomly, get the Random Scan Gibbs Sampler.}. This can be written as,
$$
\mathbf{X}^{ \left(t + 1 \right)} = \mathbf{X}^{ \left( t \right)} + r \mathbf{e},
$$
where the length $r \in \mathbb{R}$ has distribution proportional to $\pi ( \mathbf{x}^{ (t)} + r \mathbf{e} )$ \cite{liu2008monte}. 

\cite{christen2017optimal} propose as a measure of dependence, the mutual information between two random variables $\mathbf{X}$ and $\mathbf{Y}$, which measures Kullback-Leibler's divergence between the joint model $f_{\mathbf{Y},\mathbf{X}}$ and the independent alternative $f_{\mathbf{Y}}\left(\y\right)f_{\mathbf{X}}\left(\mathbf{x}\right)$, that is,
\begin{align*}
I\left(\mathbf{Y},\mathbf{X}\right)=\int\int f_{\mathbf{Y},\mathbf{X}}\left(\y,\mathbf{x}\right)\log\frac{f_{\mathbf{Y},\mathbf{X}}\left(\y,\mathbf{x}\right)}{f_{\mathbf{Y}}\left(\y\right)f_{\mathbf{X}}\left(\mathbf{x}\right)}\, d\mathbf{x}\, d\y.
\end{align*}

From the properties inherited from the Kullback-Leibler divergence $I\geq0$ and, from the Jensen inquality it is easy to prove that $I=0$ if and only if $f_{\mathbf{Y}, \mathbf{X}} = f_{\mathbf{Y}} \left(\y \right) f_{\mathbf{X}}\left(\mathbf{x}\right)$, i e., if and only if $\mathbf{X}$ and $\mathbf{Y}$ are independent.

From mutual information, \cite{christen2017optimal} explore a criterion of optimality for the Direction Gibbs algorithm. This criterion consists in minimizing the mutual information between two consecutive steps, $\mathbf{X}^{\left( t \right)}$ and $\mathbf{X}^{\left( t+1 \right)}$, of the Markov chain generated by the algorithm. They also propose, in a heuristic way, an direction distribution for the case where the target distribution is the MTN distribution. They take the directions, $\mathbf{e}$, as the eigenvectors of the precision matrix $\A$, so $\mathbf{e} = \{\mathbf{e}_1,\mathbf{e}_2, \ldots, \mathbf{e}_n \}$. The $i$-th direction will be selected with probability proportional to $\lambda_i^{-b}$, where $\lambda_i$ is the eigenvalue corresponding to the $i$-th eigenvector, $i = 1, 2, \ldots , n$, and $b$ is a random variable with $Beta$ distribution. Then, the probability of selecting the $i$-th direction is given by
\begin{equation*}
h_1 \left(\mathbf{e}_{i}\right) = \lambda_{i}^{-b}/k_1,
\end{equation*}
where $k_1 = \sum_{i=1}^{n} \lambda_{i}^{-b}$. See \cite{christen2017optimal} for more details.

Now, let $\mathbf{X} = \mathbf{X}^{\left(t\right)}$ and $\mathbf{Y} = \mathbf{X}^{ \left( t \right)} + r \mathbf{e}$ be two consecutive steps, and denote by $X_{i},Y_{i}$, $i=1,\ldots,n$, the elements of $\mathbf{X}$ and $\mathbf{Y}$, respectively. In \cite{Cricelio2016} propose to the Mutual Information as dependence measure, but now no longer on the complete vectors $\mathbf{X}$ and $\mathbf{Y}$, instead, they obtain it with $Y_{i}$ and the full vector $\mathbf{X}$, they call it marginal mutual information and is written as $I_{\mathbf{e}} \left(Y_{i}, \mathbf{X}^{\left( t \right)} \right)$, that is,
\begin{align*}
I_{\mathbf{e}} \left(Y_{i}, \mathbf{X} \right) = \int \int f_{Y_{i}, \mathbf{X}} \left(y, \mathbf{x} \right) \log\frac{f_{Y_{i},\mathbf{X}} \left(y, \mathbf{x} \right)} {f_{Y_{i}} \left(y \right)f_{\mathbf{X}} \left(\mathbf{x} \right)}\, d \mathbf{x}\, dy.
\end{align*}
The idea is to choose directions for which $I_{\mathbf{e}} \left(Y_{i}, \mathbf{X} \right)$, $\forall i =1,\ldots,n,$ is minimized. In this way, the dependency of each entry of the new generated vector $\mathbf{Y}$ with the current state $ \mathbf{X}$ is reduced.

Suppose we have a multivariate normal distribution, with precision matrix $\A_{n \times n}$ and mean vector $\MU_{n \times 1}$, but with truncated support, $x_i \in (a_i, b_i )$, $-\infty \leq a_i < b_i \leq \infty$, $i = 1, \ldots , n$. The probability density function of this MTN can be written as
$$
\pi\left(\mathbf{x}; \MU,\A,\mathbf{a},\mathbf{b}\right) = \frac{\exp \left\{- \frac{1}{2} \left(\mathbf{x} - \MU \right)^{T} \A \left( \mathbf{x} - \MU\right) \right\} }{\int_{\mathbf{a}}^{\mathbf{b}} \exp\left\{ - \frac{1}{2} \left(\mathbf{x} -\MU \right)^{T} \A \left( \mathbf{x} - \MU \right) \right\} d\mathbf{x}} \mathbbm{1}_{ (\mathbf{a},\mathbf{b})}(\x).
$$

To generate samples from the MTN distribution, in \cite{Cricelio2016} take the directions $\mathbf{e}$ as the standardized columns of the covariance matrix $\A^{-1}$, so $\mathbf{e} = \{\mathbf{e}_1,\mathbf{e}_2, \ldots, \mathbf{e}_n \}$. The $i-th$ direction will be selected with probability ($h_2 \left( \mathbf{e}_{i} \right)$) proportional to $I_{i}^{-1}$, with
\begin{equation*}
I_{i} := \sum_{j=1}^{n} I_{\mathbf{e}_i} \left(Y_{j}, \mathbf{X} \right) =- \dfrac{1}{2}\sum_{j=1}^{n} \log \left( \rho_{ij}^{2}\right), \nonumber
\end{equation*}
where $\rho_{ij}$ is the correlation between the variables $Z_{i}$ and $Z_{j}$, with $\mathbf{Z} \sim \pi$. Then, the probability of selecting the $i$-th direction is given by
\begin{equation*}
h_2 \left(\mathbf{e}_{i}\right) = I_{i}^{-1}/k_2,
\end{equation*}
where $k_2 = \sum_{i=1}^{n} I_{i}^{-1}$. See \cite{Cricelio2016} for more details of its derivation. Thus, they give more weights to the directions that make the $I_{i}$'s small.

In this article, we slightly modify the probabilities of address selection as follows.
\begin{equation*}
P_{i} := \frac{1}{n} \sum_{j=1}^{n} I_{\mathbf{e}_i} \left(Y_{j}, \mathbf{X} \right) =- \dfrac{1}{2n}\sum_{j=1}^{n} \log \left( \rho_{ij}^{2}\right), \nonumber
\end{equation*}
and the probability of selecting the $i$-th direction will be taken as
\begin{equation*}
h_2 \left(\mathbf{e}_{i}\right) = P_{i}^{-b}/k_2,
\end{equation*}
where $k_2 = \sum_{i=1}^{n} P_{i}^{-b}$ and $b$ is a random variable with $Beta$ distribution.

This article combines the algorithm given in \cite{Cricelio2016} with the modification made and the algorithm given in  \cite{christen2017optimal} for sampling MTN distributions. When the support is restricted close to the mean, the algorithm of \cite{christen2017optimal} provides a faster convergence to the target distribution, while with the algorithm given in \cite{Cricelio2016} zones of higher probability are visited. By combining both algorithms, we reduce the chain's correlations, and the support is better explored. The resulting algorithm is described in Algorithm \ref{Alg:ODG}.


\begin{algorithm}[!h]
\SetKwInOut{Input}{input}\SetKwInOut{Output}{output}
 \Input{The means vector $\MU_{n \times 1}$, the precision matrix $\A_{n \times n}$, the support $(\mathbf{a}, \mathbf{b})$, a initial value $\mathbf{X}^{ \left( 0 \right)} $, and the number of simulations $M$}
 \Output{A sample of size $M$ from $\X \sim MTN( \MU_{n \times 1}, \A_{n \times n}, \mathbf{a}, \mathbf{b})$, with the truncated support $x_i \in (a_i, b_i )$, $i = 1, \ldots , n$.}
 \BlankLine
 \textit{Step 1.} Compute the eigenvectors and eigenvalue of the precision matrix $\A$, $\{\mathbf{e}_1,\mathbf{e}_2, \ldots, \mathbf{e}_n \}$ and $\lambda_1,\lambda_2, \ldots , \lambda_n$, respectively\;
 \textit{Step 2.} Normalize the columns of the covariance matrix $\Sigma = \A^{-1}$, these will be $\{\mathbf{e}_1^c,\mathbf{e}_2^c, \ldots, \mathbf{e}_n^c \}$\;
 \textit{Step 3.} Compute the correlation matrix ($\rho$) corresponding to $\Sigma$\;
 \textit{Step 4.} Compute the weights $P_{i}$,
    $$
        P_{i} = - \dfrac{1}{2n}\sum_{j=1}^{n} \log \left( \rho_{ij}^{2}\right),
    $$
    where $\rho_{ij}$ is the correlation between the variables $X_{i}$ and $X_{j}$\;
	\BlankLine
	\For{ $t \gets 1$ \KwTo $M$ }{
	Set $\x = \X^{ \left( t -1 \right)}$\;
	Simulate $b \sim B(2., 9)$, where $B$ es the Beta distribution\;
	Simulate $p \sim U(0,1)$, where $U$ es the uniform distribution\;
	\eIf{$p < 0.5$}{
		$ h \left(\mathbf{e}_{i}\right) = \lambda_{i}^{-b}/k_1$, where $k_1 = \sum_{i=1}^{n} \lambda_{i}^{-b}$\;
		}{
		$ h \left(\mathbf{e}_{i}^c \right) = P_{i}^{-b}/k_2$, where $k_2 = \sum_{i=1}^{n} P_{i}^{-b}$\;
      }
      \textit{Step 5.} Propose a direction $ \mathbf{e} $ from the direction distribution $h(\cdot)$\;
      \textit{Step 6.} Simulate $ r \sim NT \left( \mu_{r}, \tau_{r}, c, d \right)$, where $NT$ is the univariate truncated normal distribution, $\mu_{r} = -\frac{ \mathbf{e}^{T} \A \left(\mathbf{x} - \MU \right) } { \mathbf{e}^{T} \A \mathbf{e} }$ is the mean, $\tau_{r} = \mathbf{e}^{T} \A \mathbf{e}$ is the precision, and
            \begin{align*}
            c & = \max_{i \in \left \{ 1, \ldots, n\right\} } \left( \left\{ \frac{a_{i} - x_{i}}{e_{i}} : e_{i} > 0 \right\} \cup \left\{ \frac{b_{i} - x_{i} }  {e_{i}} : e_{i} < 0 \right\} \right),\\
            d & = \min_{i \in \left\{ 1, \ldots, n\right\} } \left( \left\{ \frac{a_{i} - x_{i}}{e_{i}} : e_{i} <0 \right\} \cup \left\{  \frac{b_{i} - x_{i}}{e_{i}}: e_{i} > 0 \right\} \right)
            \end{align*}
      \textit{Step 7.} Set $\X^{ \left( t \right)} = \x + r\mathbf{e}$\;
      }
\caption{ODG: Multivariate Truncated Normal}
\label{Alg:ODG}
\end{algorithm}

\subsection{Deviance information criterion}\label{sec:Deviance}

The correlation length selection for the MTN prior model must be done carefully since each correlation length defines a different matrix $\D$, hence a different model.
Several criterions have been proposed to select between competing models.
In the maximum-likelihood framework, the most well-known criterion for model comparison is the Akaike Information Criterion (AIC), which involves the marginal likelihood  \cite{akaike1974new}. 
The Deviance Information Criterion (DIC) has been proposed as Bayesian alternative to the AIC \cite{spiegelhalter2002bayesian} to select the model that better fits the data between a pool of competing 
models.
The DIC is particularly useful in Bayesian model selection problems where the model's posterior distributions have been obtained by MCMC simulation. The DIC's advantages is that it reduces each model to a single number summary and that the models to be compared do not need to be nested.

For a model with parameters $\theta$ and data $y$, the DIC is calculated as
\begin{equation*}
DIC = \overline{D(\theta)}  + p_{D} = D(\bar{\theta}) + 2p_{D}.
\end{equation*}
where $D(\theta) := -2 \log( likelihood )= -2 \log( p(y|\theta) )$ is called the model deviance, 
$\overline{D(\theta)} = E\left[ D(\theta) | y \right]$ is the posterior expected deviance, $\bar{\theta} = E\left[ \theta | y \right]$
is the posterior mean, and $p_{D} = \overline{D(\theta)} - D(\bar{\theta})$ is called the effective number of parameters.
In our case, the expectations may be easily calculated using the MCMC sample.  We then proceed to calculate the DIC for all models of interest and choose the one with the smallest DIC value. For further discussion of the DIC see \cite{spiegelhalter2002bayesian}.


\section{Results}

\subsection{Synthetic example}\label{sec:SynCase}
To validate our solutions against a known slip movement, we generate a synthetic data set based on the same fault geometry and geodetic stations configuration as for the 2006 Guerrero SSE. For this, we assume a priori that the slips ($\D$) have a MTN with zero mean vector (i.e., in average we consider that there is no slipping), restricting the support according to the information we have on the GGap. That is, $
\D \sim N\left( \mathbf{0}, \frac{1}{\sigma_{\beta}^2} \A_{0} \right),
$ subject to $-.0804 \leq d_{d}^i \leq 0.4$,  $-0.1 \leq d_{s}^i \leq 0.1$, $i = 1,2, \ldots, M$, where $\A_{0} = \BETA \W \C^{-1} \W \BETA$, $\C$ is the correlation matrix given by Eq.~(\ref{eq:Corr}), and $\W$ is the matrix of weights computed with Eq. (\ref{eq:Wei}). For the remaining parameters we consider $\lambda_s = \laot$, $\lambda_d = \latt $, $ \sigma_{\beta}^2 = 0.0002$, $\beta_{s} = 1$ and $\beta_{d} = 0.2$. To simulate a slip we fix a displacement in a subfault $\D_i$ (approximately in the same place where the maximum displacement is suspected in the real 2006 GGap SSE) and the rest of the subfualt displacements are simulated from the conditional distribution $\D_{-i}|\D_i = d_i$ (also a MTN). 

Once the displacements vector $\D$ has been generated, we solve the forward problem ($\mathbf{U} = \XD$) adding Gaussian noise to obtain our synthetic observations. Figure \ref{fig:MAPsynt}(a) shows the true slip movement $\D$ and the synthetic measurements. Following this strategy of simulation of the synthetic data and considering the statistical model given in Sec.~\ref{sec:FM}, we obtain simulations of the MTN distribution using the ODG algorithm. We set the hyperparameter $\sigma^2_{\beta}$ as explained in appendix \ref{Ape:Sigmas}.
\begin{figure}
\centering
\includegraphics[width=.6\textwidth]{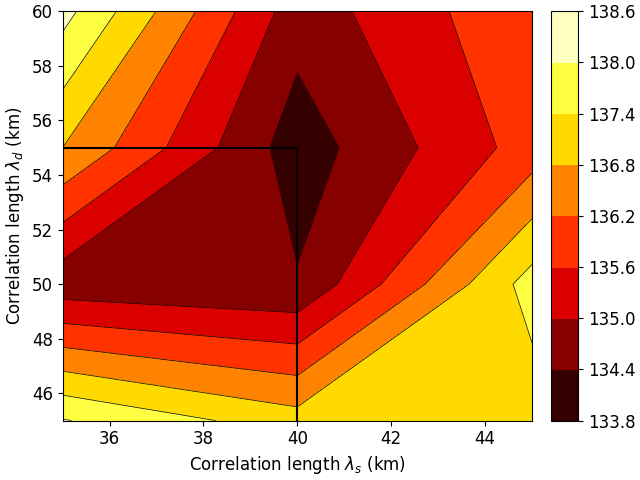}
\caption{The DIC's obtained with the MCMC output considering synthetic data, for different combinations of correlation lengths. The best value is chosen for correlation lengths of $\lambda_s = \lao$ and $\lambda_d = \lat$ km, although the other DIC's are similar. The actual correlation lengths used for simulation are $\lambda_s = \laot$ and $\lambda_d = \latt$. The Bayesian inversion is quite similar in all cases (see Figure~\ref{fig:MAPsynt}), as predicted by the similarities in the DIC's. The selection of the correlation length in this scale is robust and, on the other hand, in general estimating correlation is commonly a difficult statistical problem.}
\label{fig:DICssim}
\end{figure}

We consider different correlation lengths for $\lambda_s$ and for $\lambda_d$ in the a priori distributions, see Fig.~\ref{fig:MAPsynt} and~\ref{fig:Unc_syn}. For the optimal correlation lengths, we computed the DIC in a grid search along the hyperparameter space $\lambda_s:[35,45]\textrm{ km}\times \lambda_d:[45,60]$ km, see Fig.~\ref{fig:DICssim}. The optimal correlation lengths of $\lambda_s = \lao$  and $\lambda_d = \lat$ were finally selected.

\subsection{Posterior distribution and uncertainty representations}
\begin{figure}
\subfigure[True synthetic displacements]{\includegraphics[scale=0.5]{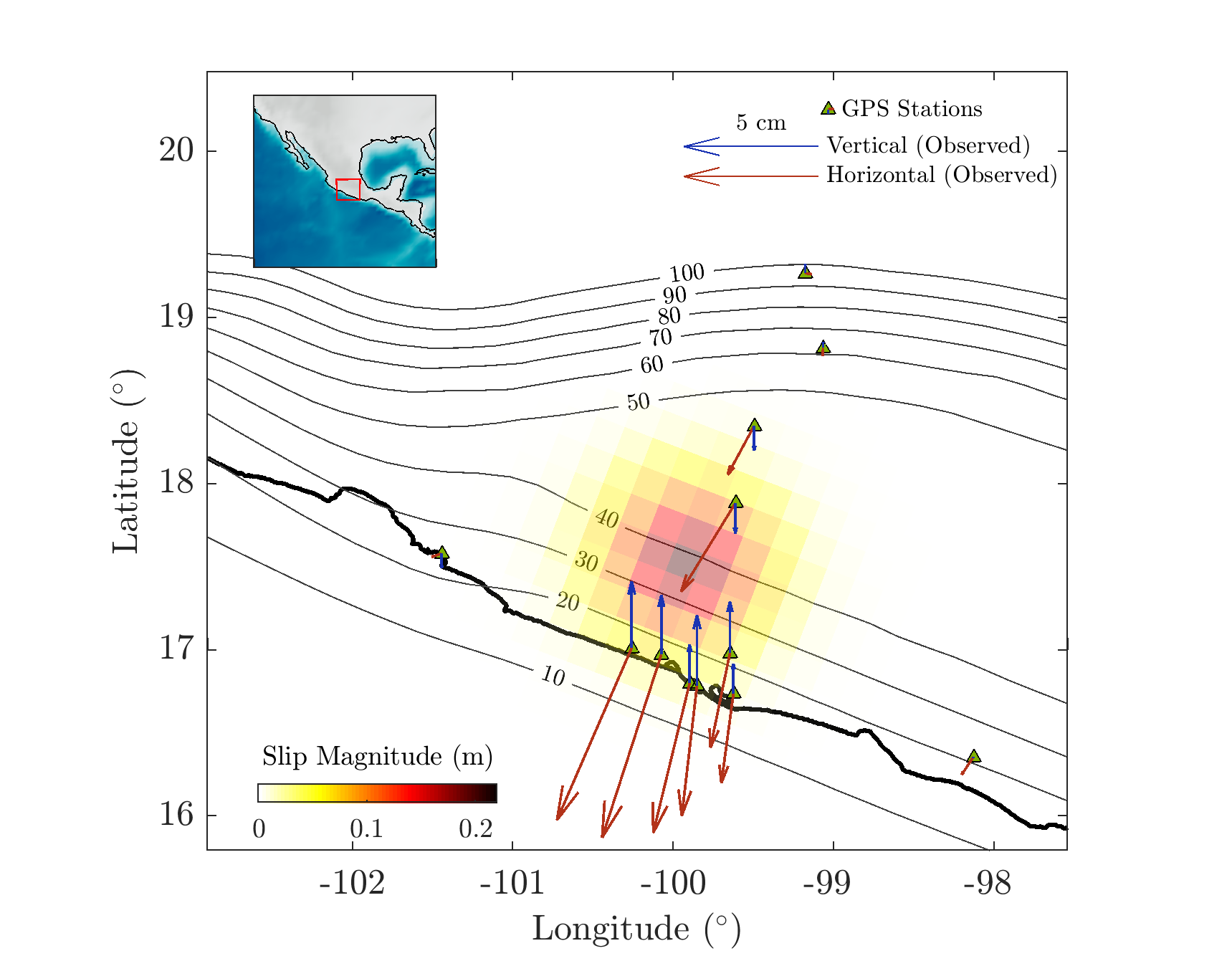}}
\subfigure[Bayesian inversion: $\lambda_s = 35, \lambda_d=50$]{\includegraphics[scale=0.5] {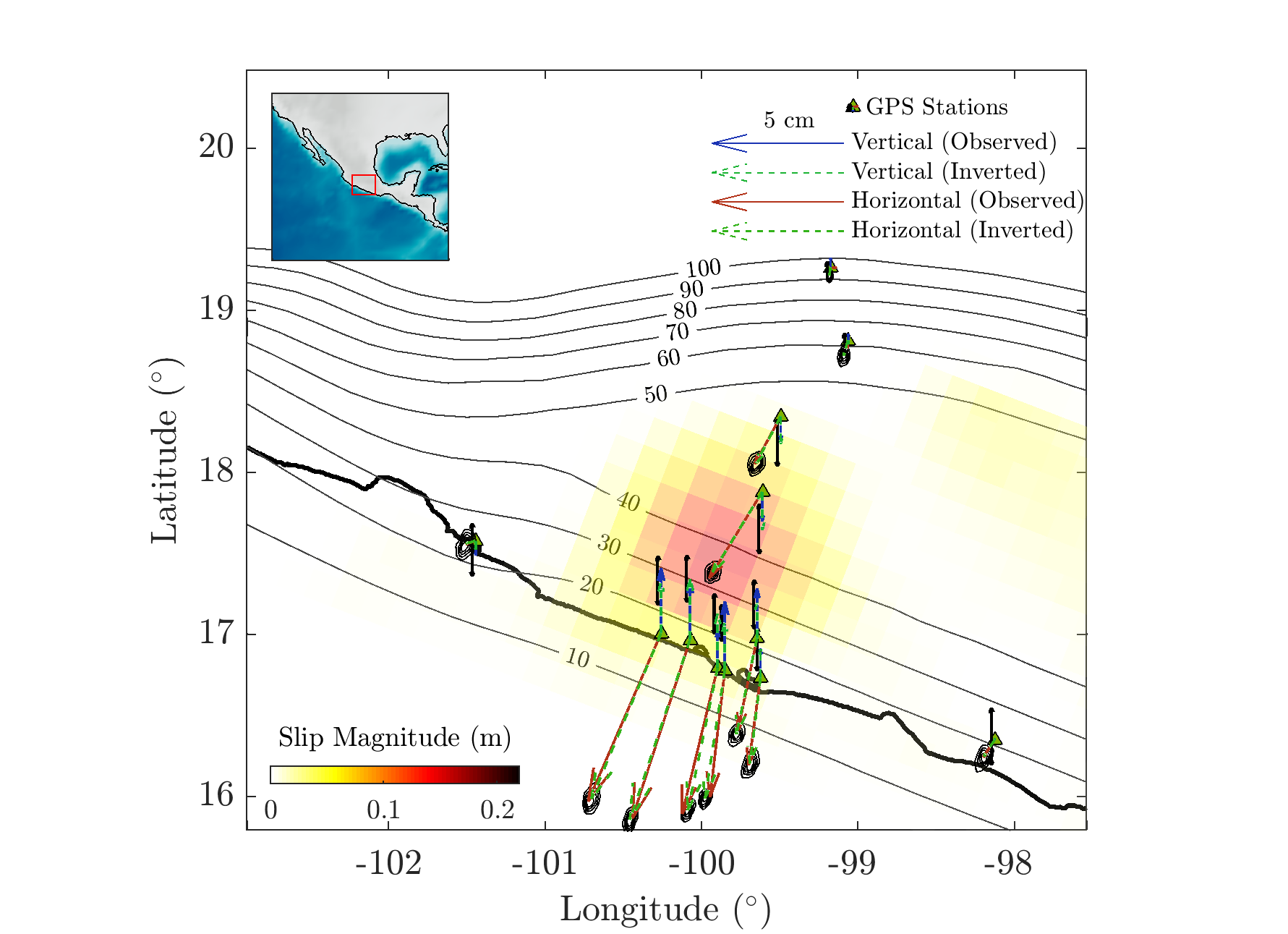}}
\subfigure[Bayesian inversion: $\lambda_s = 40, \lambda_d = 45$]{\includegraphics[scale=0.5]{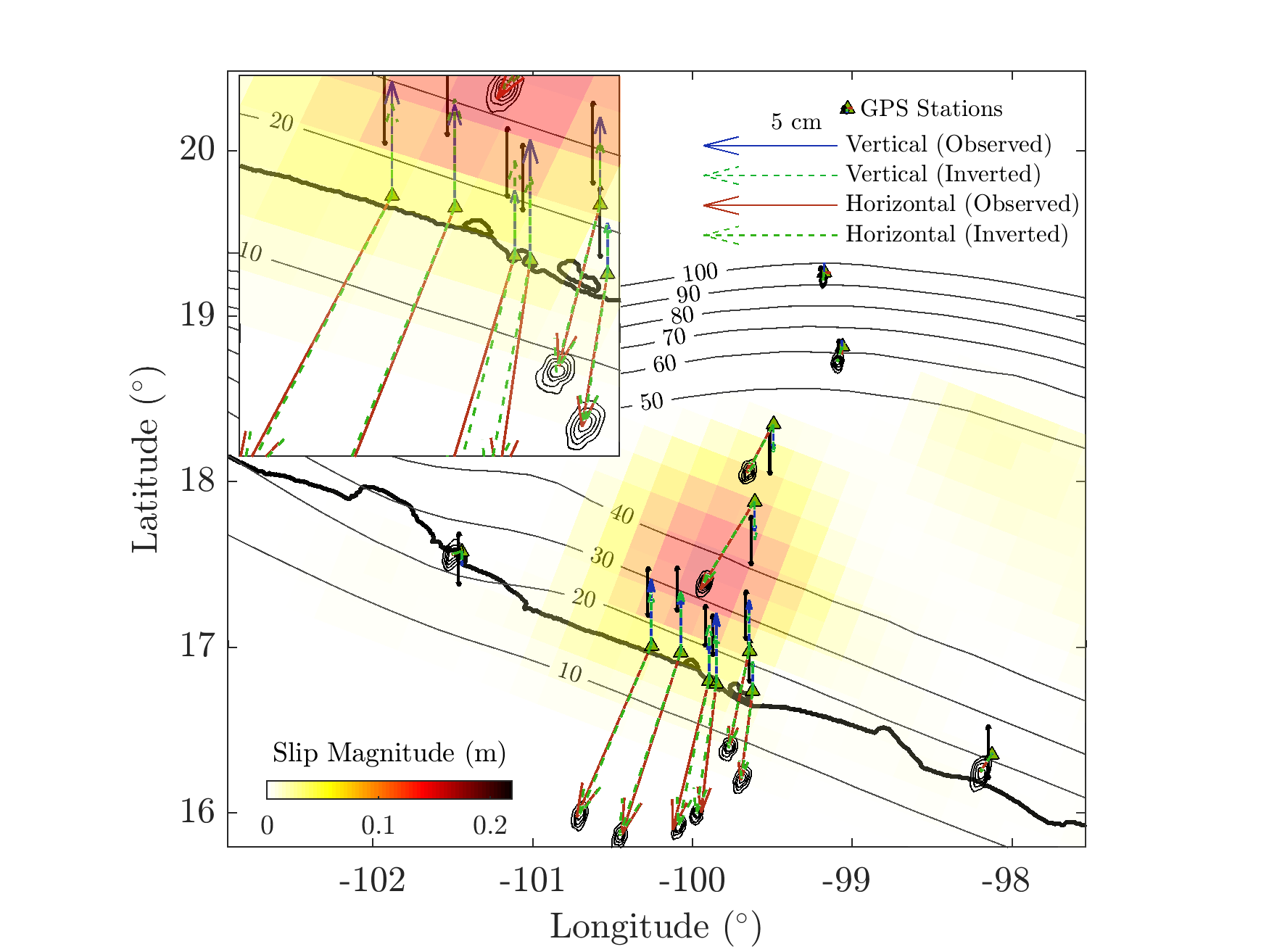}} 
\subfigure[MLE]{\includegraphics[scale=0.5]{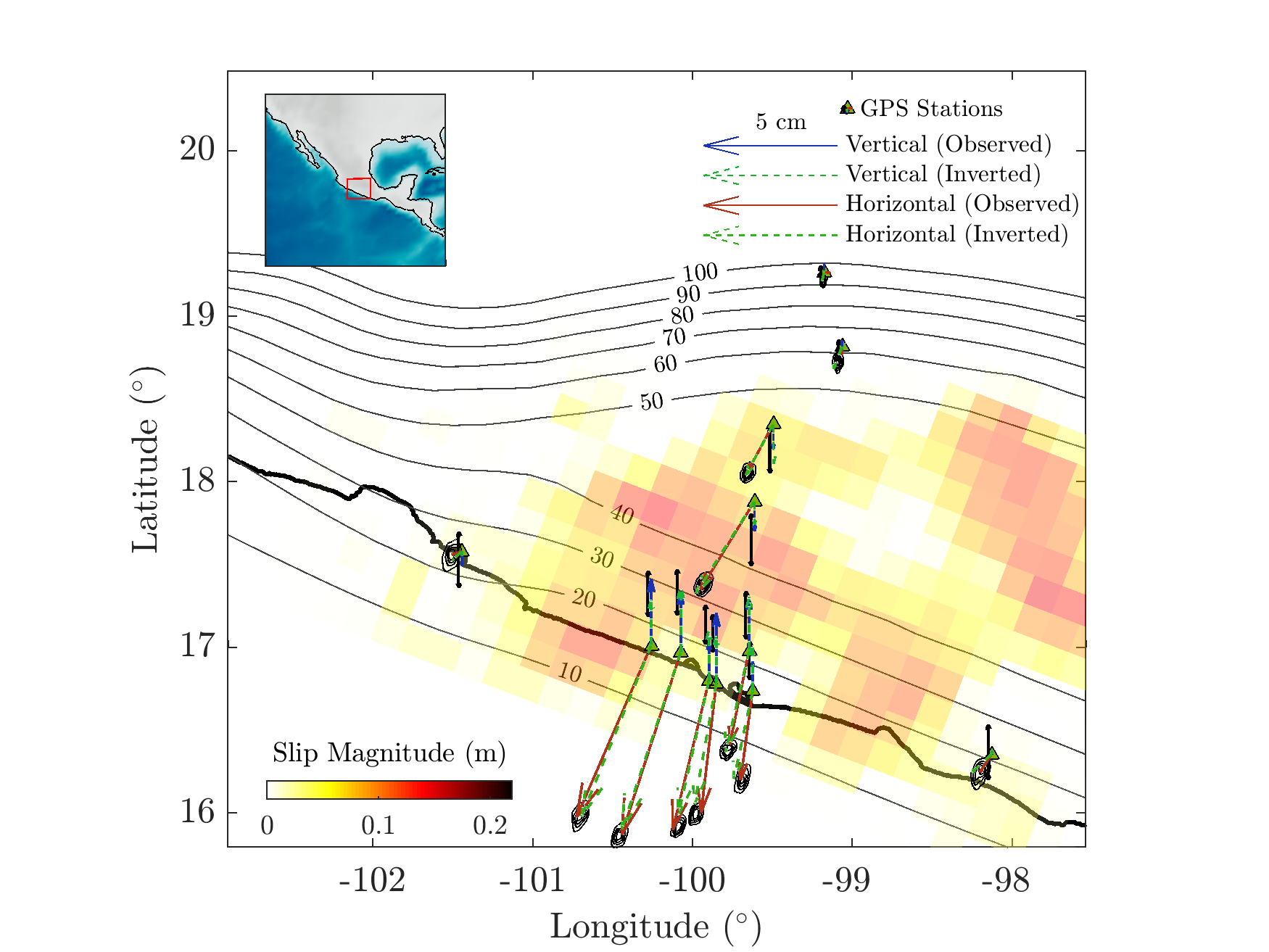}}
\caption{Slip models on the plate interface (background colors) and the associated model displacement predictions (arrows) with different correlations lengths: (a) True synthetic displecements, (b) and (c) Bayesian inversion using correlation lengths  $\lambda_s = 35, \lambda_d = 50$ and $\lambda_s = 40, \lambda_d = 45$ repectively . (d)  MLE estimator.
Blue-solid and red-solid arrows show the observed (synthetic) surface displacements while dashed arrows show the predictions. The black contours at the arrowheads represent the data uncertainty; see the zoomed insert in (c). The median of the posterior samples for the slip inversion are shown in (b) and (c) while the MLE is shown in (d) (heat colors). Green triangles show the GPS station locations.  Black lines represent the isodepth contours (in km) of the subducted oceanic slab.}
\label{fig:MAPsynt}
\end{figure}

The median of the posterior samples of each subfault was plotted for the chosen correlation lengths in Figs.~ \ref{fig:MAPsynt} (b) and (c), with a heat map.  We also plot the GPS stations locations with triangles and the their corresponding data using arrows.  The black contours at the arrowheads represent the posterior error calculated for the inversion in the data. 
We can see that the data fit is excellent in all cases and the slip solution is almost perfect, in relation to the true slip seen in Fig.~ \ref{fig:MAPsynt} (a).

Since we have access to the full posterior distribution, we can look at point estimators such as the posterior mean, the posterior median, the maximum a posteriori (MAP). However, these point estimators may be unrepresentative of the actual posterior. The mean and median may be misleading for long-tailed asymmetric PDFs, and the MAP may be unrepresentative in the presence of skewness. In this synthetic case, we obtain far better results with the median of the posterior samples, shown in Figs.~\ref{fig:MAPsynt}(b)--(c). For comparisons, and to observe the crucial importance of the inclusion of correctly modeled prior information, we include a Maximum Likelihood Estimation in Fig.~\ref{fig:MAPsynt} (d). 

\begin{figure}
\subfigure[Bayesian inversion: $\lambda_1 = 40, \lambda_2 = 55$]{\includegraphics[scale=0.8]{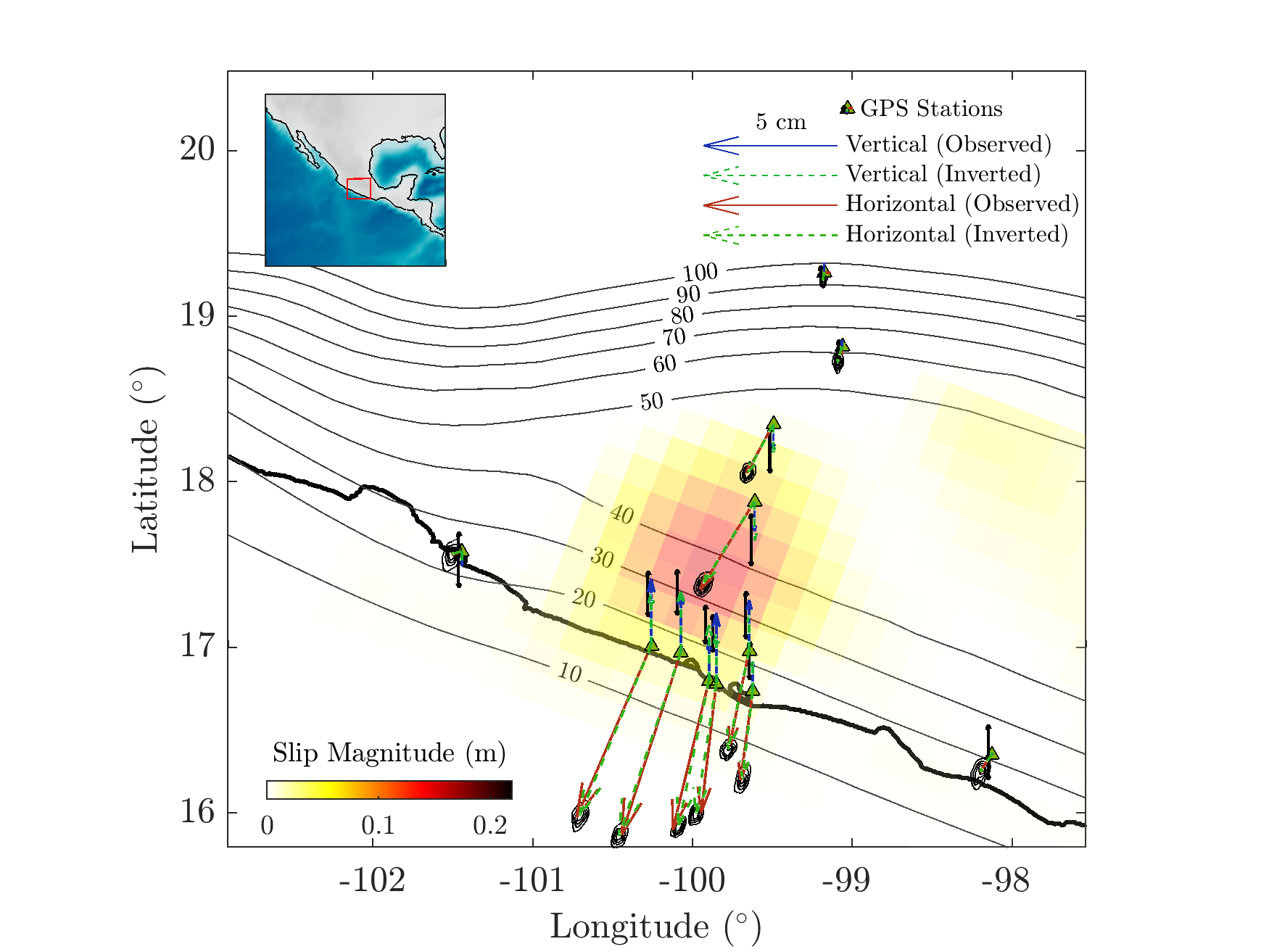} }
\subfigure[Coefficient of variation]{\includegraphics[scale=0.8]{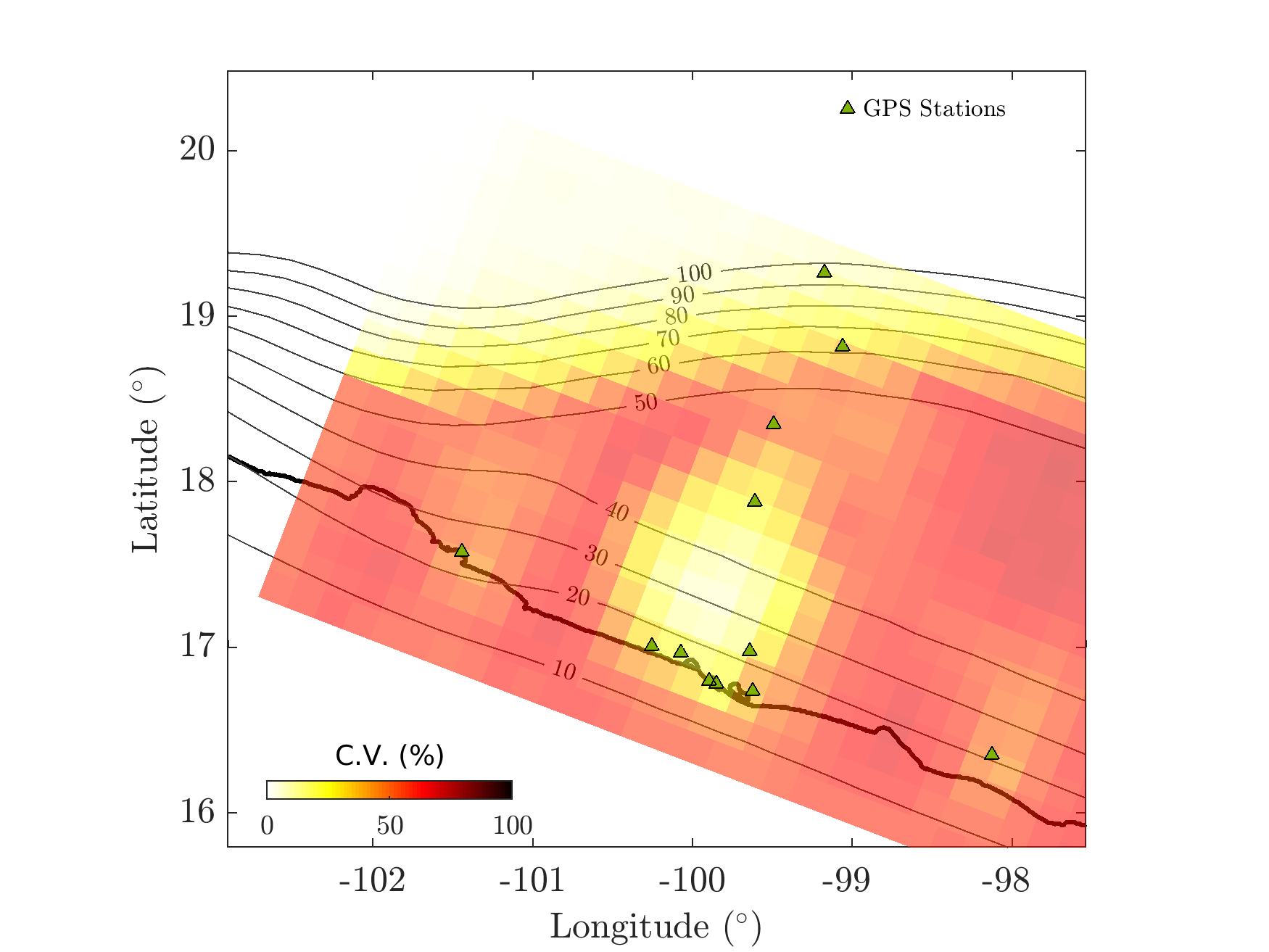}} 
\caption{Slip posterior median (a) and its coefficient of variation (CV) (b), resulting from the Bayesian inversion of synthetic GPS cumulative displacements for the optimal correlation lengths $\lambda_s = \lao$ and $\lambda_d = \lat$, see Fig.~\ref{fig:MAPsynt} for more details. The CV is expressed as a percentage and quantifies the uncertainty in the inverted slip shown in (a).  Higher (darker) CV values implies more uncertainty in the inferred slip.}
\label{fig:Unc_syn}
\end{figure}

An advantage of the Bayesian approach is that it does not only produce one optimal model, but the sampling yields a large ensemble of probable models, sampled from the posterior distribution. In Figs.~\ref{fig:Unc_syn} (a) and (b), we represent the median and the uncertainty of the slips considering $\lambda_s = \lao$  and $\lambda_d = \lat$, respectively. These correlation lengths provide the lowest DIC (see Fig.~\ref{fig:DICssim}).

Now we turn to the question of uncertainty representation. This is not a straight forward task for a posterior distribution on a vector field. Here we resort to the Coefficient Variation (CV). Which the CV is a statistical measure of the dispersion of probability distribution around its mean. The CV represents the ratio of the standard deviation $\sigma$ to the mean $\mu$ (CV = $\frac{\sigma}{\mu}$), showing a relative quantity of the degree of variation, independent of the scale of the variable. This metric provides a tool to compare the data dispersion between different data series. We use the MCMC simulations to estimate the CV at each subfault. In Fig.~\ref{fig:Unc_syn} (b) the CV is plotted to compare the posterior uncertainty in the inferred (inverted) movements in each subfault.  We can see that in the areas where the largest slip was found, we have the least relative uncertainty, that is, the median is more representative. Note also that where the GPS stations are located, we have a clear decrease in uncertainty. These regions of low uncertainty are consistent with the regions with maximum restitution index computed by \cite{Tago_2020} through a mobile checkerboard strategy. The map with the displacement medians along with the map with their corresponding CVs, as in Fig.~\ref{fig:Unc_syn}, is our representation of the posterior distribution, and the UQ representation of this inversion.  
In the next section we use the same strategy to study the 2006 GGap SSE inversion.

\section{Real case: 2006 Guerrero Slow Slip Event}\label{sec:Sismo}
In this section we present a real data application to illustrate the performance of our approach. We study the 2006 Guerrero SSE with data collected by the Instituto de Geof\'isica (IGF), Universidad Nacional Autónoma de M\'exico (UNAM), and the Servicio Sismol\'ogico Nacional (SSN). In 2006, a SSE in Guerrero was recorded by $N = 15$ GPS stations. The stations are located mainly along the coast and on a transect perpendicular to the trench, between Acapulco and the north of Mexico city \cite{radiguet2011spatial}.  We used these same locations in the synthetic analysis presented in the previous Section.

\subsection{Observations and data preprocessing}
Regarding the observations, we assume that some small number of GPS stations are available on the surface.  The GPS data must be preprocessed taking into account the time-varying climate phenomena. Besides, the inter-SSE steaty-state motion is subtracted to isolate the GPS data related with an SSE event, that is, the tectonic coupling is removed. For the actual GPS data in the GGap 2006 event, we used the data processed by \cite{radiguet2011spatial} with the their proposed standard deviations, $\sigma_{x} = 0.0021$, $\sigma_{y} = 0.0025$, and $\sigma_{z} = 0.0051$ in the north, east and vertical directions, respectively. All these quantities are measured in meters. The time window that was considered to compute the displacements was from January 2, 2006 to May 15, 2007.

We solve the Bayesian inversion as a multiple linear regression model with constraints on the coefficients, considering the statistical model \eqref{eq:Model} as explained in Sect.~\ref{sec:FM}. For the GPS data, we took $\sigma_{x}^{2} = 0.0021$, $\sigma_{y}^{2} = 0.0025$, and $\sigma_{z}^{2} = 0.0051$ as the standard deviations in the north, east and vertical directions, respectively \cite{radiguet2011spatial}. As in the synthetic case, the regularization parameter,\stcris{s, $\sigma^2$ and} $\sigma^2_{\beta}$, is obtained by minimizing Eq.~\eqref{Eq:Variances} given in Appendix \ref{Ape:Sigmas}. The time window used to compute the displacements was from January 2, 2006 to May 15, 2007.

\begin{figure}
\centering
\includegraphics[width=.6\textwidth]{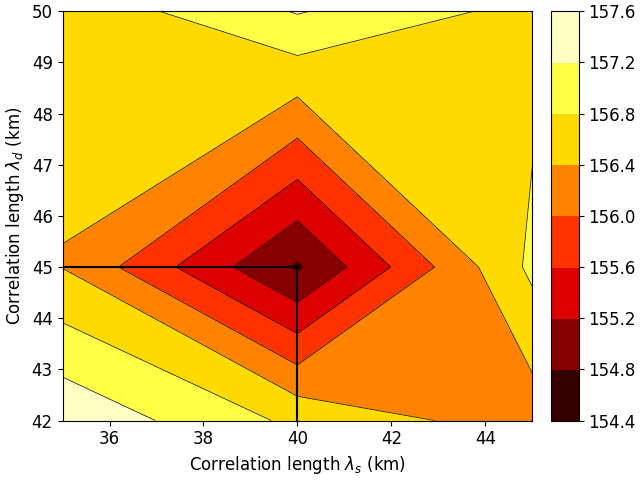}
\caption{Results for the 2006 Guerrero SSE Bayesian inversions: The DICs obtained with the MCMC output, for different combinations of correlation lengths. The best value is chosen for a correlation lengths of $\lambda_s = 40$ km  and $\lambda_d = 45$ km.}
\label{fig:DICsreal}
\end{figure}

To sample from the resulting MTN posterior distribution we use the OGD sampler explained in Algorithm \ref{Alg:ODG}. For the optimal correlation lengths, we computed the DIC in a grid search along the hyperparameter space $\lambda_s:[30,45]\textrm{ km}\times \lambda_d:[42,50]$ km, see Fig.~\ref{fig:DICsreal}. The optimal correlation lengths of $\lambda_s = 40$  and $\lambda_d = 45$ were finally selected.

\begin{figure}
\subfigure[Bayesian inversion: $\lambda_s = 40, \lambda_d = 45$]{\includegraphics[scale=.78]{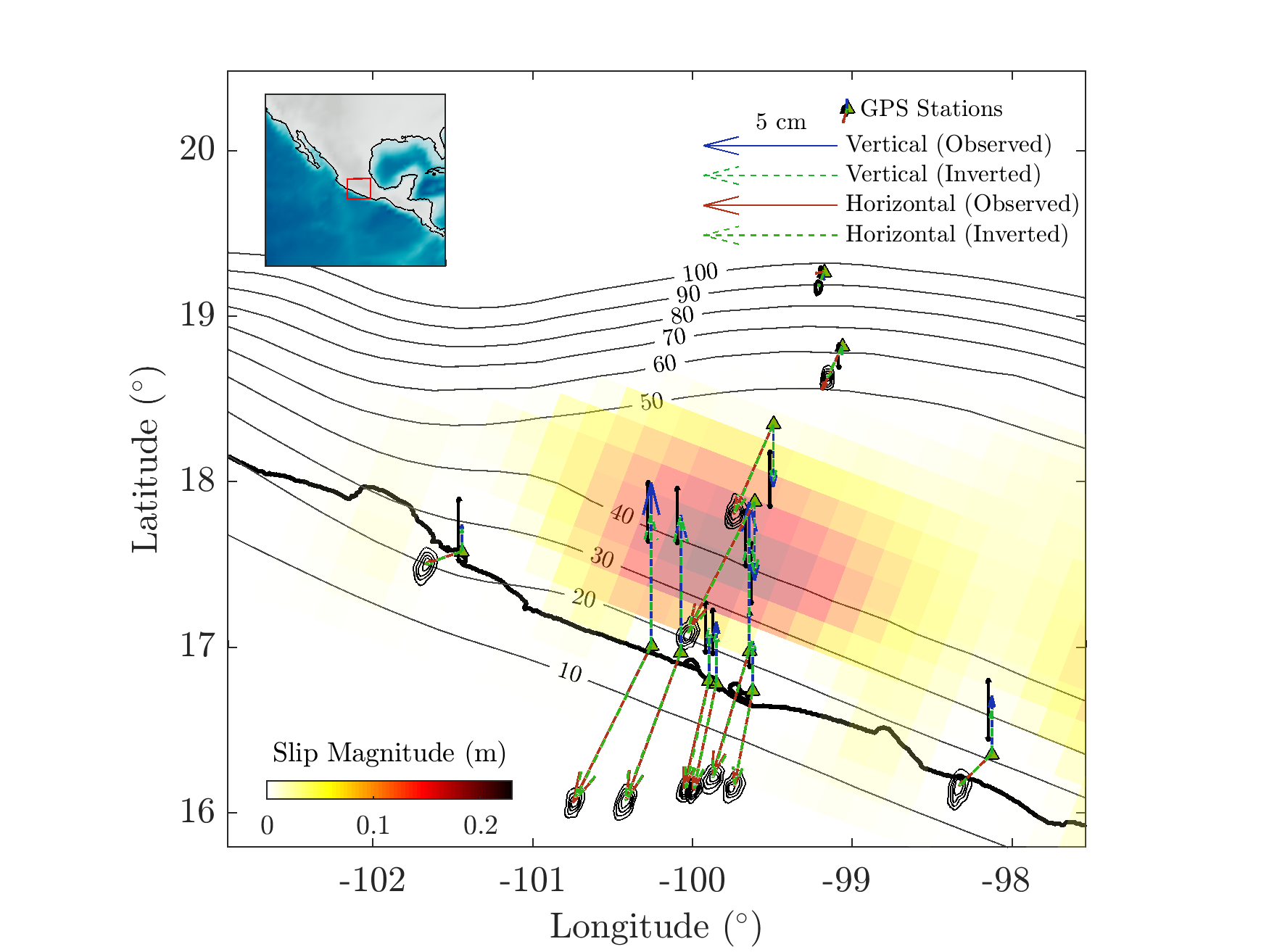}}
\subfigure[Coefficient of variation]{\includegraphics[scale=0.78]{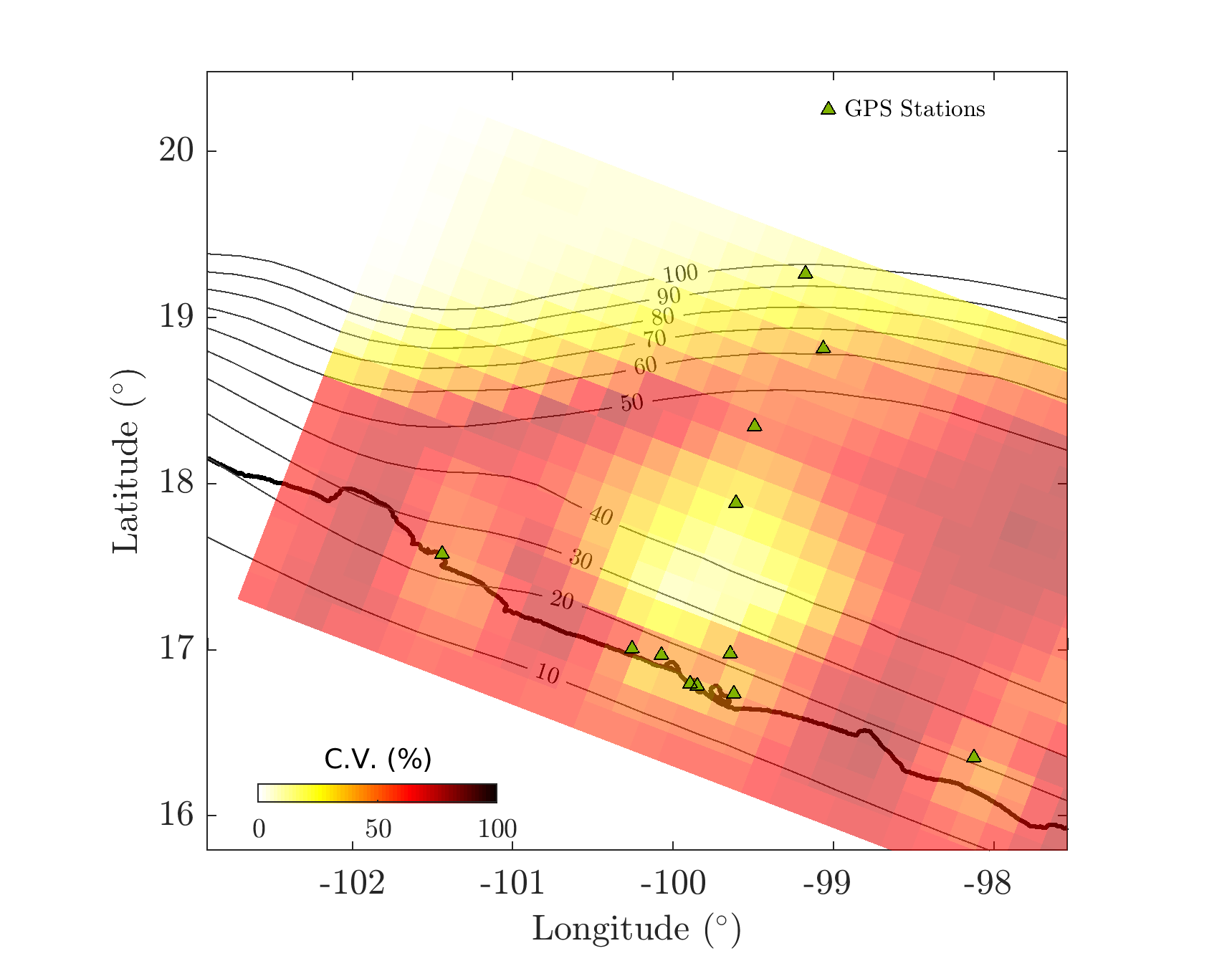}}
\caption{Bayesian inversion of fault displacements of the 2006 Guerrero SSE for correlation lengths $\lambda_s = 40, \lambda_d = 45$, following the same conventions as in Fig.~\ref{fig:MAPsynt}. (a) Posterior median of the slip and (b) its CV, expressed as a percentage.}
\label{fig:Real_results}
\end{figure}

The median of the posterior samples for the static inversion was plotted (heat colors) in Fig.~\ref{fig:Real_results}. The black contours at the arrowheads represent the data uncertainty, and the horizontal lines in the vertical component represent the quantiles $0.025, 0.5, 0.975$, respectively.  The CV is plotted in Fig.~\ref{fig:Real_results} (b).

All GPS data is well recovered by the method within the estimated uncertainty bounds. The median of the posterior show 
a compact region where most of the displacement took place. It is consistent with the most recent inversions, where the region of maximum slip is located from 30 to 40 km depth and with a slight updip penetration in the north-west section \cite{Bekaert_2015,Tago_2020}. Recent offshore observations showed that the mechanical properties in that segment of the subduction slab are different and it may explain the inferred updip slip \cite{Plata-Martinez_2020}. Despite the similarities, it is important to mention that most of previous works are supported on constrained optimization framework and their solutions corresponds to the MAP, a different point-wise estimate than the one presented here (e.g., \cite{radiguet2011spatial}; \cite{Tago_2020}). As explained in the introduction, these estimates may be biased and the comparison with our results should be made carefully. 

A novelty of our procedure is depicted in Fig.~\ref{fig:Real_results} (b) where we are able to estimate the posterior CV, as measure of the uncertainty in our solution. 
Uncertainty is low in the region where the fault's displacement is concentrated, and nearby
coastline GPS stations.  While the former is a consequence of solution to the IP,
the latter is expected since the GPS station illuminates the nearby faults. 
On the upper part of the color map, where the Cocos plate dives into the mantle, the uncertainty is also low. 
This is a consequence of the prior information built into our prior distribution.  Specifically, by using the weight matrix
$\W$, and represents our knowledge that the Cocos and North American plates are not coupled at such depths.  Therefore, very low or no uncertainty in the displacement is to be observed in this area. 

\subsection{Uncertainty Quantification of the moment magnitude}
Given a particular displacements vector $\D$, the moment magnitude $M_w(\D)$ is computed as
$$
M_w(\D) = \frac{2}{3} \left(\log _{10} \frac {M_{0}(\D)}{N \cdot m } - 9.1 \right),
$$
where $M_{0}(\D)  = \mu A \D$ is the seismic moment in N$\cdot$m, $\mu$ is the crustal rigidity in $Pa$ and $A$ is the surface that slipped in $m^2$ \cite{stein2009introduction}. We take the $1$ cm slip contour as the effective SSE area, and we consider a typical crustal rigidity $\mu = 32x10^9$ Pa.

\begin{figure}
\subfigure[Synthetic]{\includegraphics[scale=0.5]{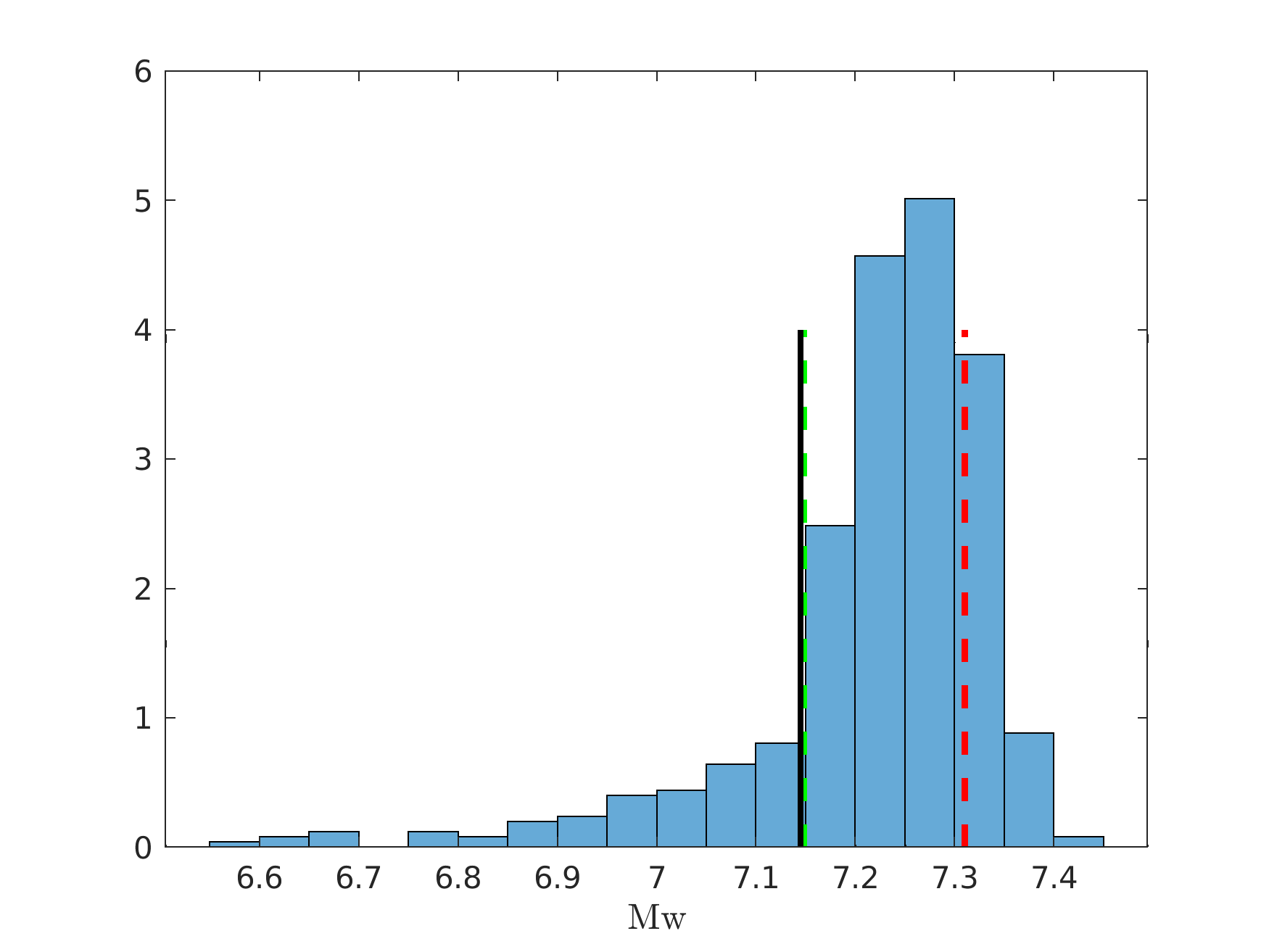}} 
\subfigure[2006 Guerrero SSE]{\includegraphics[scale=0.5]{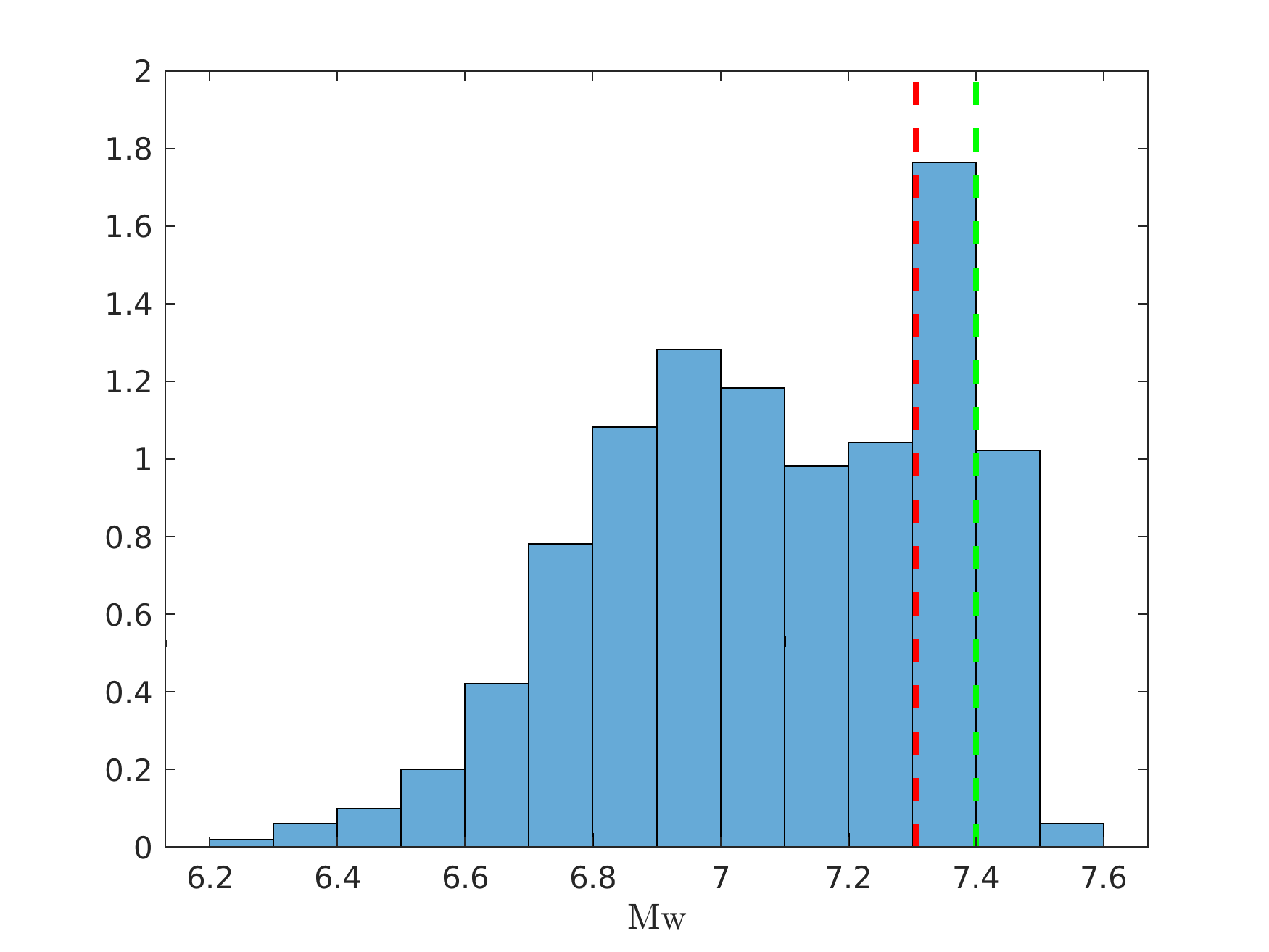}} 
\caption{Posterior distributions of the moment magnitude $M_w$, along with the MAP (red) and median (green) point estimates: (a) Synthetic case fault displacement inversion presented in Fig.~\ref{fig:Unc_syn}, the black marker represents the true $M_w$. (b) 2006 Guerrero SSE fault displacement inversion presented in Fig.~\ref{fig:Real_results}.}
\label{fig:mw}
\end{figure}

A further advantage of the Bayesian approach is that it may consistently produce estimates and UQ of inferred parameters. That is, the posterior distribution of $M_w(\D)$ is well defined, as the transformation of the random vector $\D | \mathbf{Y}$.  Moreover, since we already have a Monte Carlo sample, $\D^{(1)}, \D^{(2)}, \ldots , \D^{(T)}$, of the posterior $\D | \mathbf{Y}$, $M_w^{(i)} = M_w(\D^{(i)})$, $i = 1, 2, \ldots,T$, is a MC sample from the posterior distribution of the moment magnitude. The posterior distributions for the Mw of the synthetic and 2006 Guerrero SSE examples are presented in Fig.~\ref{fig:mw}.
For comparisons, also the moment magnitude of the MAP and the median displacements are plotted in Fig.~\ref{fig:mw}.
In Fig.~\ref{fig:mw} (a), the synthetic case, the true $M_w$ is plotted; indeed calculated from the true displacements used to simulate the exercise, seen in Fig.~\ref{fig:MAPsynt} (a). Note how the posterior of $M_w$ is skewed and would suggest larger values that what is estimated using the MAP or the median. The $M_w$ of the median coincides nicely with the true value and both are contained in the posterior (note that for this non-linear functional, the $M_w$ of the median displacements need not to coincide with the median of the posterior for $M_w$).\\
For the 2006 Guerrero SSE, we can extract from the posterior distribution of $M_w$ the point estimates $M_{w_\mathrm{median}}=7.4$ and $M_{w_\mathrm{MAP}}=7.3$. The former is consistent with the the value computed by \cite{Tago_2020}, the later with study by \cite{Bekaert_2015} and both are below the $M_w=7.5$ estimated by \cite{radiguet2011spatial}. Clearly our point estimates are consistent with previous studies, however the skewed posterior distribution has not been showed before. From the synthetic data, it seems that the median is a better point estimate but further research must be done.

\section{Discussion} \label{sec:Dis}

Solutions to IPs that include UQ in geophysics are challenging problems. Computationally feasible methods such as Tikhonov regularization may introduce biases due to non-physically justified regularization terms and solutions may misrepresent part of the phenomena. Moreover, different considerations in the regularization terms may produce dissimilar solutions.  Bayesian methods provide a natural alternative to explore the full posterior distribution of IPs and provide formal UQ.


In this work, we developed an efficient Bayesian approach to estimate fault slips in a 
SSE. We calculated the full fault slip posterior through a constrained multiple linear regression 
model and geodetic observations. We postulate a Gaussian model for geodetic data, 
and a MTN prior distribution for the unknown slip. The resulting fault slip posterior pdf is also a MTN. 
Regarding the posterior, we proposed  an efficient Optimal Direction Gibbs sampler algorithm to sample from this high-dimensional MTN; it can be carried out in a personal computer as opposed  to other MCMC samplers that require high-performance computing. An advantage of our  algorithm is that no parameter has to be adapted or tuned.

Prior elicitation is a fundamental part of the modeling  process regarding the particular physical 
problem under study. We use the Mat\'ern covariance function to control the subfault autocovariance   
and to impose physically-consistent slip restrictions (on the prior distribution). 
Different correlation lengths are considered in the prior distribution. As a model selection criteria, we propose the DIC to choose an optimal correlation length. 


In many applications the MAP estimator is chosen as a representative solution to the IP due to its  computational feasibility 
by usual regularization schemes.  Our results in the synthetic data case shows that the MAP is biased. 
Instead, we propose  the posterior median as an alternative to fault slips's point-wise estimate. 
Moreover, UQ is represented by the coefficient of variation. We compare variability 
between subfaults and show areas where we have the most certainty.  Since we have Monte Carlo samples of the
full posterior distribution, both these quantities are readily available, in sharp contrast to regularization methods, where these quantities can not be recovered. 
For the 2006 Guerrero SSE, the median of the posterior distribution of the slip shows a compact slip patch where most of the slip is located from 30 to 40 km depth with a slight updip penetration in the north-west section. Both of these main characteristics are consistent with the most recent studies of \cite{Bekaert_2015} and \cite{Tago_2020}. Besides those coincidences, through the CV we can assess the uncertainty which is lower where the most of the slip is located. With the posterior distribution of the slip, we could easily compute the posterior distribution of $M_w$ which showed to be skewed with the median and MAP estimates also consistent with the above mentioned studies.


One limitation of our current approach is that the FM needs to be linearized in order to obtain a MTN posterior.  A more general FM could also be analyzed, although adding severe computational burden and difficulties in a highly multidimensional MCMC.     


The Bayesian framework allows us to also consider different representations of uncertainty. It is clear that more SSE's should be analyzed for the GGap, and elsewhere. A further improvement would be to learn parameters of the particular fault slip from multiple SSE's analyses such as the correlation length. Moreover, border effects should be formally included in the covariance matrix, by improving the covariance operator \cite{daon2018mitigating}. For the moment, border effects do not seem apparent in the maps produced by the median, as seen in the examples presented here.  However, we leave this ideas for future research.


Computationally efficient Bayesian methods are being developed for many IPs in geophysics. 
In many cases, they provide access to full posterior distributions, which provide better and 
more informative estimates for the solutions as well as UQ.
Our proposed methodology is applied to a real data set, for the 2006 Guerrero SSE, 
where the objective was to recover the slip on a known interface from observations at few geodetic stations. However, our method can be used for any earthquake slip inversion; as long as the FM can be linearized. Lastly, once the slip inversion is available, calculating the seismic moment, with formal UQ, is a simple subproduct of our methodology, a result that, up to our knowledge, is new. 

\section*{Acknowledgments}
JCML, AC and JAC are partially founded by CONACyT grants CB-2016-01-284451 and COVID19-312772 and a RDECOMM grant. AC was also partially supported by UNAM PAPPIT–IN106118 grant. JT was also partially supported by CONACyT grant 255308.
\newpage
\bibliographystyle{unsrt}
\bibliography{bibliography}

\newpage
\appendix
\section{Determining the variances $\sigma^{2}$ and $\sigma_{\beta}^{2}$} \label{Ape:Sigmas}

Akaike's Bayesian Information Criterion (ABIC), proposed by \cite{akaike1980likelihood}, has been widely applied in geophysical inversion to determine the regularization parameters $\left(\sigma^{2},\sigma_{\beta}^{2}\right)$. Following ABIC, we propose a criterion by maximizing the marginal posterior distribution of these parameters, that is,
\begin{align}
\max : f \left( \sigma^{2}, \sigma_{\beta}^{2}| \Y \right) & =\int f \left( \D, \sigma^{2}, \sigma_{\beta}^{2} | \Y \right) d \D = \frac{1}{f \left( \Y \right)} \int f \left( \Y, \D, \sigma^{2}, \sigma_{\beta}^{2} \right) d \D. \label{eq:MargPos}
\end{align}
For this, we consider the follow hierarchical linear model
\begin{align}
\Y| \D,\sigma^{2} & \sim N \left(\XD, \frac{1}{\sigma^{2}} \A \right), \label{eq:HierMod}\\
\D|\sigma_{\beta}^{2} & \sim N_{m}\left( \boldsymbol{0}, \frac{1}{\sigma_{\beta}^{2}} \A_{0}\right); \quad\A_{0} = \BETA \W \C^{-1} \W \BETA, \nonumber \\
\sigma_{\beta}^{2} & \sim\text{Inv-Gamma}\left(a_{\beta}, b_{\beta}\right), \nonumber \\
\sigma^{2} & \sim\text{Inv-Gamma}\left(a,b\right),\nonumber 
\end{align}
where $\sigma^{2}\A =\boldsymbol{\Sigma}^{-1}$ is the precision
matrix, and $\text{Inv-Gamma}\left(\alpha,\beta\right)$ denote a
inverse Gamma distribution with shape parameter $\alpha$ and scale
parameter $\beta$. Note that,
\begin{align*}
\int f\left(\Y,\D,\sigma^{2},\sigma_{\beta}^{2}\right)d\boldsymbol{\beta} & =\int f\left(\Y,\D|\sigma^{2},\sigma_{\beta}^{2}\right)\pi\left(\sigma_{\beta}^{2}\right)\pi\left(\sigma^{2}\right)d\D\\
& =\pi\left(\sigma_{\beta}^{2}\right)\pi\left(\sigma^{2}\right)\int f\left(\Y,\D|\sigma^{2},\sigma_{\beta}^{2}\right)d\D\\
& =\pi\left(\sigma_{\beta}^{2}\right)\pi\left(\sigma^{2}\right)m\left(\Y|\sigma^{2},\sigma_{\beta}^{2}\right),
\end{align*}
where $\pi\left(\sigma_{\beta}^{2}\right)$ and $\pi\left(\sigma^{2}\right)$
are the prior distributions for $\sigma_{\beta}^{2}$ and $\sigma^{2}$,
respectively, and
\begin{equation}
m \left( \Y| \sigma^{2}, \sigma_{\beta}^{2} \right) := \int f\left(\Y,\D|\sigma^{2},\sigma_{\beta}^{2}\right)d\D.\label{eq:Marginal}
\end{equation}
So, maximizing (\ref{eq:MargPos}) is equivalent to minimizing
\[
\min: \ell \left( \sigma^{2}, \sigma_{\beta}^{2} \right) = -\log \left(m \left( \Y| \sigma^{2}, \sigma_{\beta}^{2} \right)\right)-\log\left(\pi\left(\sigma_{\beta}^{2}\right)\right)-\log\left(\pi\left(\sigma^{2}\right)\right).
\]
Now, with the hierarchical model (\ref{eq:HierMod}) and using the derivation of (\ref{eq:Marginal}) given in \cite{xu2019akaike}, that is,
\begin{align*}
m \left( \Y| \sigma^{2}, \sigma_{\beta}^{2}\right) & = \frac{1}{\left( 2\pi\right)^{n/2}\sqrt{ \left| \Sigma_{py} \right|}} \exp \left\{ -\frac{1}{2} \left( \Y - \X\MU_{0} \right)^{T} \Sigma_{py}^{-1} \left( \Y - \X\MU_{0}\right) \right\} ,
\end{align*}
where $\MU_{0} = \boldsymbol{0}$ is the prior mean of
$\D$, and $\Sigma_{py}=\A^{-1}\sigma^{2}+\X\A_{0}^{-1}\X^{T}\sigma_{\beta}^{2}$, it follows that, the optimal variances are obtained by minimizing
\begin{equation*}\label{Eq:Variances}
\min:\ell\left(\sigma^{2}, \sigma_{\beta}^{2}\right) = \ln\left\{ \det \left( \Sigma_{py} \right) \right\} +\Y^{T} \Sigma_{py}^{-1} \Y + \left( a_{\beta}+1 \right) \log \left( \sigma_{\beta}^{2} \right) + \frac{b_{\beta}}{\sigma_{\beta}^{2}} + \left( a + 1 \right) \log \left( \sigma^{2} \right) + \frac{b}{\sigma^{2}}. 
\end{equation*}
If $\sigma^{2}$ is given/known but $\sigma_{\beta}^{2}$ unknown, the optimal prior variance $\sigma_{\beta}^{2}$ is obtained by minimizing 
\begin{equation*}\label{Eq:Variance}
\min: \ell \left( \sigma_{\beta}^{2} \right) = \ln \left\{ \det \left( \Sigma_{py} \right) \right\} +\Y^{T} \Sigma_{py}^{-1} \Y + \left( a_{\beta}+1 \right) \log \left( \sigma_{\beta}^{2} \right) + \frac{b_{\beta}}{\sigma_{\beta}^{2}}. 
\end{equation*}

Minimizing the above expressions is straightforward since are functions defined in $\mathbb{R}^2$ and $\mathbb{R}^1$, respectively.

\end{document}